%% file: main.tex
\documentclass[journal,a4paper]{IEEEtran}

\usepackage{xcolor}
\usepackage{balance}
\usepackage{cite}
\usepackage{amsmath,amssymb,amsfonts}
\usepackage{graphicx}
\usepackage{textcomp}
\usepackage{acronym}
\usepackage{xcolor}
\usepackage{tikz} 
\usepackage[utf8]{inputenc}
\usepackage{pgfplots} 
\usepackage{pgfgantt}
\usepackage{pdflscape}
\usepackage{changes}
\usepackage{comment}
\usepackage{subfigure}
\usepackage{microtype}
\usepackage{mathtools,algpseudocode,algorithm,MnSymbol}
\usepackage{geometry}
\usepackage{multirow}

\usepackage{pgfplots}
  \pgfplotsset{compat=newest}
  \usetikzlibrary{plotmarks}
  \usetikzlibrary{arrows.meta}
  \usepgfplotslibrary{patchplots}
  \usepackage{grffile}
  \usepackage{amsmath}
  
\newcommand{\beginsupplement}{%
        \setcounter{table}{0}
        \renewcommand{\thetable}{S\arabic{table}}%
        \setcounter{figure}{0}
        \renewcommand{\thefigure}{S\arabic{figure}}%
       \setcounter{equation}{0} \def\theequation{S\arabic{equation}}
     }

\pgfplotsset{compat=newest} 
\pgfplotsset{plot coordinates/math parser=false} 

\allowdisplaybreaks
\input{acronyms.tex}

\usepackage{amsthm}

\newtheorem{example}{Example}

\newtheorem{lemma}{Lemma}

\newtheorem{theorem}{Theorem}

\begin{document}

\bibliographystyle{IEEEtran}
\bstctlcite{IEEEexample:BSTcontrol}
\title{Batch SLAM with PMBM Data Association Sampling and Graph-Based Optimization}
\author{Yu Ge,~\IEEEmembership{Member,~IEEE,}   
Ossi Kaltiokallio,
Yuxuan Xia,~\IEEEmembership{Member,~IEEE,}\\
\'Angel F. Garc\'ia-Fern\'andez,
Hyowon Kim,~\IEEEmembership{Member,~IEEE,} 
Jukka Talvitie,~\IEEEmembership{Member,~IEEE,} \\
Mikko Valkama,~\IEEEmembership{Fellow,~IEEE,}
Henk Wymeersch,~\IEEEmembership{Fellow,~IEEE,}
Lennart Svensson,~\IEEEmembership{Senior~Member,~IEEE} 
\thanks{Yu Ge,  Henk Wymeersch, and Lennart Svensson are with the Department of Electrical Engineering, Chalmers University of Technology, Gothenburg, Sweden. Emails: 
\{yuge,henkw,lennart.svensson\}@chalmers.se.}     
\thanks{Ossi Kaltiokallio, Jukka Talvitie, and Mikko Valkama are with the Unit of Electrical Engineering, Tampere University, Tampere, Finland. Emails: \{ossi.kaltiokallio,jukka.talvitie,mikko.valkama\}@tuni.fi.} 
\thanks{ Yuxuan Xia is with the Department of Electrical Engineering, Link\"oping University, Link\"oping, Sweden. Email: yuxuan.xia@liu.se.}
\thanks{ \'Angel F. Garc\'ia-Fern\'andez is with the Department of Electrical Engineering and Electronics, University of Liverpool, Liverpool, United Kingdom. Email: Angel.Garcia-Fernandez@liverpool.ac.uk.}
\thanks{Hyowon Kim is with the Department of Electronics Engineering, Chungnam National University, Daejeon, South Korea. Email: hyowon.kim@cnu.ac.kr.}
\thanks{This work was partially supported by the Wallenberg AI, Autonomous Systems and Software Program (WASP) funded by Knut and Alice Wallenberg Foundation, the Vinnova Beyond5GPOS project, and the Research Council of Finland (RCF) under the grants \#345654, \#352754, \#357730, and \#359095.}
% ...

}



\maketitle

\begin{abstract}
\Ac{SLAM} methods need to both solve the \ac{DA} problem and the joint estimation of the sensor trajectory and the map, conditioned on a \ac{DA}. In this paper, we propose a novel integrated approach to solve both the DA problem and the batch SLAM problem simultaneously, combining \ac{RFS} theory and the graph-based SLAM approach. A sampling method based on the \ac{PMBM} density is designed for dealing with the \ac{DA} uncertainty, and a graph-based SLAM solver is applied for the conditional SLAM problem. In the end, a post-processing approach is applied to merge SLAM results from different iterations. Using synthetic data, it is demonstrated that the proposed SLAM approach achieves performance close to the posterior Cram{\'e}r-Rao bound, and outperforms state-of-the-art RFS-based SLAM filters in high clutter and high process noise scenarios. 

\end{abstract}

\vskip0.5\baselineskip

\acresetall

\section{Introduction}
The objective of the \acf{SLAM} problem \cite{ThrunProb2005, durrant2006simultaneous} is to deduce the dynamic pose of a mobile sensor over time, along with constructing a map of the surrounding environment, using measurements obtained from one or multiple sensors. Drawing inspiration from pioneering research in autonomous robotics \cite{smith1990estimating}, the \ac{SLAM} problem has captured broad interest in recent decades: it holds significant importance with a multitude of applications spanning diverse fields, such as robotics \cite{ThrunProb2005}, autonomous driving \cite{bresson2017simultaneous}, virtual and augmented reality \cite{wang2016comprehensive}, indoor navigation \cite{dardari2015indoor,yassin2016recent}, integrated sensing and communication \cite{di2014location,ge2023mmwave}, and so on. 

Traditional SLAM methods typically follow a two-step approach: i) solve the \ac{DA} problem between the unknown number of landmarks and imperfect measurements, which may include clutter and mis-detections, ii) estimate the joint posterior density of the sensor trajectory and the map, given measurements, control inputs, and the DA from step (i). Two important methodologies are the \emph{filtering-based} and \emph{graph-based} approaches. In filtering-based approaches \cite{durrant2006simultaneous,smith1990estimating,dissanayake2001solution}, the map is modeled with an unknown number of physical landmarks with unknown spatial locations, and the map and the sensor state are then typically estimated sequentially.  Prominent examples are \ac{EKF}-\ac{SLAM} \cite{dissanayake2001solution} and FastSLAM \cite{montemerlo2002}, which has been demonstrated to work well in practice, but is sensitive to \ac{DA} uncertainty \cite{neira2001}. 

On the other hand, in graph-based approaches \cite{dellaert2006square,thrun2006graph,grisetti2010tutorial}, the sensor state at a specific time step or a physical landmark is represented as a node in a graph, and each edge represents probabilistic dependency between two sensor states, or between a landmark state and a sensor state.  The sensor trajectory and the map can be simultaneously estimated by obtaining the \ac{MAP} estimate, optimizing over the whole graph. Unlike filtering-based approaches, graph-based SLAM typically takes all measurements and performs optimization techniques on the entire graph, maintaining cross-correlation information between the sensor trajectory and the map. This results in more robust and accurate estimates, and makes graph-based SLAM perform batch processing and typically work offline. Among graph-based SLAM approaches, the GraphSLAM algorithm has become a prevalent offline \ac{SLAM} solver for batch processing, due to its global consistency properties \cite{falchetti2017random}. However, the performance of graph-based SLAM heavily relies on the quality of the DA. Statistical tests such as the $\chi^{2}$  test, joint compatibility test, or other types of heuristics are often applied to solve the \ac{DA} problem \cite{neira2001data}, which could fail in complex scenarios.

One theoretically appealing approach to handling DAs is using \acp{RFS} \cite{mahler2014advances}. Modeling the map and measurements as \acp{RFS}  enables a fully integrated Bayesian SLAM solution that treats the \ac{DA} uncertainty as a part of the estimation process \cite{mullane2011random}. In \ac{RFS}-based SLAM frameworks, different \acp{RFS} are used to model the map, resulting in \ac{PHD}-SLAM filters in \cite{mullane2011random,kim20205g,EKPHD2021Ossi}, the \ac{LMB}-SLAM filters in \cite{deusch2015labeled,deusch2016random}, the \ac{GLMB}-SLAM filters in \cite{moratuwage2018delta,moratuwage2019delta},  the \ac{PMBM}-SLAM filters in \cite{ge20205GSLAM,ge2022computationally}, and the \ac{PMB}-SLAM filters in \cite{ge2022iterated,kim2024setBP}. Within these RFS-based SLAM solutions, the \ac{PMBM}-SLAM filters can explicitly consider all possible \acp{DA},  resulting in better performance by sacrificing time efficiency. 
Many RFS-based SLAM solutions, such as \cite{kim20205g,ge2020exploiting,ge20205GSLAM}, apply \ac{rbp} filter, similar as the FastSLAM solution, sampling the sensor trajectory and taking \ac{RFS} likelihoods into consideration in the particle weight computation. 
To reduce the computational complexity, the number of required particles can be reduced by using  an approximation of the \ac{OID}
 to draw samples efficiently \cite{kaltiokallio2022towards,kaltiokallio2023multi}. In addition, low complexity alternatives are introduced in  \cite{ge2022computationally,EKPHD2021Ossi,ge2022iterated,leitinger2019belief,kim2024setBP}, which rely on linearization and jointly updating the sensor state and the map, dropping the cross-correlation between the sensor and the map, while keeping the \ac{RFS} format. These approaches can have relatively low computational complexities by sacrificing the \ac{SLAM} performance and robustness. 

Although batch solutions to the backend problem are known to yield superior performance, all existing RFS-based SLAM solutions focus on filtering. Considering that the \ac{DA} problem in batch SLAM resembles the \ac{DA} problem in the batch \ac{MTT} problem, DA techniques designed for \ac{MTT} can be leveraged to address the DA in graph-based SLAM. One possible solution to the \ac{DA} association problem in \ac{MTT} is to use sampling-based methods \cite{granstrom2017likelihood,fatemi2017poisson,casella1992explaining}, which have been shown to exhibit excellent performance in challenging scenarios. \ac{MCMC} sampling  methods were proposed in \cite{fatemi2017poisson,xia2023efficient,granstrom2017likelihood} to handle the \ac{DA} problem, using the Gibbs sampling \cite{casella1992explaining} or/and the merge-split \ac{MH} algorithms \cite{jain2004split}.

In this paper, we present the first method that combines the advantages of batch processing with RFS for a theoretically optimal treatment of the DAs. The proposed approach can overcome the limitation of RFS-based SLAM methods, which are restricted to sequential processing, and graph-based SLAM methods, which rely on heuristics to handle the \ac{DA} problem. Our approach iteratively applies two methods: (i) an \ac{MCMC} sampling method based on the PMBM density to solve the \ac{DA} uncertainty;  (ii) a graph-based SLAM solver for a set of landmarks and the sensor trajectory conditioned on a specific  \ac{DA} and existences of landmarks. In the end, the final sensor trajectory and the map are acquired through a post-processing marginalization step, which involves merging the SLAM results from different iterations and considering the undetected landmarks. Our main contributions are summarized as follows:
\begin{itemize}
    \item \textbf{The development of a novel SLAM algorithm}: We designed a new Graph PMBM-SLAM approach, which embodies a cyclic process of sampling, and graph-based SLAM. 
    The framework bridges RFS theory and graph-based SLAM, where the RFS theory is leveraged to devise a sampling-based method for addressing the \ac{DA} uncertainty, and graph-based SLAM serves as an optimal solution for tackling the SLAM problem given a determined \ac{DA} and existences of landmarks. This integration provides a new effective and robust SLAM solution. Via simulation, this iterative refinement process achieves performance close to the \ac{PCRB}, along with high accuracy and robustness in challenging scenarios. 
    \item \textbf{The derivation of a new \ac{MCMC} sampling method for batch SLAM}: Based on the RFS theory, a novel \ac{MCMC} sampling method is formulated for addressing the \ac{DA} problem. The proposed sampling method combines the Gibbs and the \ac{MH} algorithms and exhibits superior performance compared to the Gibbs sampling and the \ac{MH} algorithms on their own, providing reliable \ac{DA} solutions for  the batch SLAM problem. 
    \item \textbf{The derivation of a novel marginalization algorithm for post-processing}: The GraphSLAM algorithm tackles the SLAM problem given a determined \ac{DA} and existences of landmarks. By merging GraphSLAM results for different posterior samples of \ac{DA} and existences of landmarks, and considering the undetected landmarks, the sensor trajectory and the PMB representation of the set of landmarks are estimated, including the existence probability of each detected landmark. 
  
\end{itemize}

The subsequent sections of this article are structured as follows: Section \ref{Sec:system} details the system models and introduces the fundamental concepts of the \ac{PMBM} density. Section \ref{Sec:framework} focuses on the proposed Graph PMBM-SLAM approach designed for batch processing. Section \ref{data_associatio_section} elaborates on the representation of the \ac{DA} problem and its solution using a sample-based approach. In Section \ref{Sec:GraphSLAM_givenDA}, the exploration is directed towards the GraphSLAM algorithm, conditioned on a specific \ac{DA} sample and existences of landmarks. Section \ref{Sec:fuse_over_samples} delineates the method for merging the map and sensor trajectory across iterations.
The detailed demonstration of the simulated environment and the presentation of simulation results are provided in Section \ref{Sec:results}.
Finally, Section \ref{Sec:conclusion} summarizes our concluding remarks.

\subsubsection*{Notations}
Scalars (e.g., $x$) are denoted in italic, vectors (e.g., $\boldsymbol{x}$) in bold, matrices (e.g., $\boldsymbol{X}$) in bold capital letters, sets  (e.g., $\mathcal{X}$) in calligraphic. The cardinality of a set or the number of elements in a sequence of sets is denoted by $\left|\,\cdot\,\right|$. The inner product of $f(\boldsymbol{x})$ and $g(\boldsymbol{x})$ is denoted by $\left\langle f;g \right\rangle=\int f(\boldsymbol{x}) g(\boldsymbol{x}) \mathrm{d}\boldsymbol{x}$. The transpose is denoted by $(\cdot)^{\mathsf{T}}$, and the union of mutually disjoint sets is denoted by $\biguplus$. A multivariate Gaussian distribution with mean $\boldsymbol{u}$ and covariance $\boldsymbol{\Sigma}$ is denoted as $\mathcal{N}(\boldsymbol{u},\boldsymbol{\Sigma})$, and $d_{\boldsymbol{x}}=\text{dim}(\boldsymbol{x})$ is the dimension of $\boldsymbol{x}$. The $i$-th component of $\boldsymbol{x}$  is denoted by $[\boldsymbol{x}]_{i}$, and the component in the $i$-th row and  $j$-th column of $\boldsymbol{X}$  is denoted by $[\boldsymbol{X}]_{i,j}$.
\vspace{-3mm}
\section{Models and PMBM Background} \label{Sec:system}
\vspace{-1mm}
\subsection{Sensor, Landmark, and Measurement Models}
The sensor state at time step $k$, denoted as $\boldsymbol{s}_{k}$, includes various components depending on the specific problem and scenario. The transition density of $\boldsymbol{s}_{k}$ can be expressed as \cite{durrantwhyte2006}
\begin{equation}
f(\boldsymbol{s}_k | \boldsymbol{s}_{k-1}) = {\cal N}(\boldsymbol{s}_k ; \boldsymbol{v}(\boldsymbol{s}_{k-1}),\boldsymbol{Q}_{k-1}), \label{transition_density}
\end{equation}
where $\boldsymbol{v}(\cdot)$ denotes a known transition function, and $\boldsymbol{Q}_{k-1}$ denotes a known covariance matrix.
The map of the environment consists of various landmarks, and we model the map as an \ac{RFS}, denoted as $\mathcal{X} = \lbrace \boldsymbol{x}^{1}, \; \ldots, \; \boldsymbol{x}^{I} \rbrace$, where $I=|\mathcal{X}|$ represents the total number of landmarks, and each element $\boldsymbol{x}^{i} \in\mathcal{X}$ denotes a specific landmark state. It is worth noting that both $I$ and $\boldsymbol{x}^{i} \in \mathcal{X}$  are random, as $\mathcal{X}$ is modeled as an \ac{RFS} \cite[Section 2.3]{mahler2014advances}.

We assume a point object model, where each landmark can generate at most one measurement per time instant. The detection probability $p_{\text{D}}(\boldsymbol{x}^{i},\boldsymbol{s}_{k}) \in [0,1]$ is introduced to account for how likely there is a measurement from landmark $\boldsymbol{x}^{i}$, when the sensor has state $\boldsymbol{s}_{k}$. At time step $k$, a set of measurements $\mathcal{Z}_{k}=\{\boldsymbol{z}_{k}^{1},\dots, \boldsymbol{z}_{k}^{\hat{{I}}_{k}} \}$ is observed, where ${\hat{{I}}_{k}}$ is the number of measurements, and  $\boldsymbol{z}_{k}^{i} \in \mathcal{Z}_{k}$ is a specific measurement. Assuming measurement  $\boldsymbol{z}^{i}$ has originated from landmark $\boldsymbol{x}^{i}$, its likelihood is given by
\begin{align}
    f(\boldsymbol{z}_{k}^{i}|\boldsymbol{x}^{i},\boldsymbol{s}_{k})=\mathcal{N}(\boldsymbol{z}_{k}^{i};\boldsymbol{h}(\boldsymbol{x}^{i},\boldsymbol{s}_{k}),\boldsymbol{R}_k^i),\label{pos_to_channelestimation}
\end{align}
where $\boldsymbol{h}(\boldsymbol{s}_{k},\boldsymbol{x}^i )$ is the known measurement function, which is a function of the sensor state and the landmark state, and $\boldsymbol{R}_k^i$ is the corresponding covariance matrix.
It is important to note that usually $\hat{{I}}_{k} \neq {{I}}_{k}$, is due to clutter and missed detections. Apart from landmark-generated measurements, $\mathcal{Z}_{k}$ may contain clutter that is modeled as a \ac{PPP} (see \eqref{PPP}), parameterized by the intensity function $c(\boldsymbol{z})$.

\subsection{PMBM Density}
Suppose a map is modeled as an \ac{RFS} $\mathcal{X} = \lbrace \boldsymbol{x}^{1}, \; \ldots, \; \boldsymbol{x}^{I} \rbrace$, which is characterized by the set density $f(\mathcal{X})$. In the \ac{PMBM} representation of the map,  the set of landmarks is separated into two disjoint sets: the set of undetected landmarks, which are the landmarks that have never been detected, and the set of detected landmarks, which are the landmarks that have been detected at least once. Therefore, $\mathcal{X}$ can be divided into two mutually disjoint sets. 

The set of undetected landmarks $\mathcal{X}_{\mathrm{U}}$ is modeled as a \ac{PPP}, and the set of detected landmarks is modeled as an \ac{MBM} $\mathcal{X}_{\mathrm{D}}$, which results in $\mathcal{X}=\mathcal{X}_\mathrm{U} \biguplus \mathcal{X}_\mathrm{D}$ following a \ac{PMBM} density \cite{williams2015marginal,garcia2018poisson,fatemi2017poisson}. The \ac{PPP} density $f_{\mathrm{P}}(\mathcal{X}_{\mathrm{U}})$ is given by
\begin{equation}
    f_{\mathrm{P}}(\mathcal{X}_{\mathrm{U}})=e^{-\int\lambda(\boldsymbol{x})\mathrm{d}\boldsymbol{x}}\prod_{\boldsymbol{x} \in \mathcal{X}_{\mathrm{U}} }\lambda(\boldsymbol{x}),\label{PPP}
\end{equation}
where $\lambda(\cdot)$ is the intensity function, and the density can be parameterized by $\lambda(\boldsymbol{x})$. The \ac{MBM} density $f_{\mathrm{MBM}}(\mathcal{X}_{\mathrm{D}})$ is 
\begin{equation}
    f_{\mathrm{MBM}}(\mathcal{X}_{\mathrm{D}})= \sum_{j \in \mathbb{J}}w^{j}\sum_{\biguplus_{i \in \mathbb{I}^{j}}  \mathcal{X}^{i}=\mathcal{X}_{\mathrm{D}}}\prod_{i=1}^{|\mathbb{I}^{j}|}f^{j,i}_{\mathrm{B}}(\mathcal{X}^{i}),\label{MBM}
\end{equation}
where  $\mathbb{J}$ is the index set of all global hypotheses, which corresponds to \acp{DA} in the \ac{SLAM} problem ~\cite{williams2015marginal},  $w^{j}\ge 0$ is the weight for $j$-th global hypothesis, satisfying $\sum_{j\in\mathbb{J}}w^{j}=1$, and $\mathbb{I}^{j}$ is the index set of landmarks (i.e., the Bernoulli components) under the $j$-th global hypothesis with density
\begin{equation}
f^{j,i}_{\mathrm{B}}(\mathcal{X}^{i})=
\begin{cases}
1-r^{j,i} \quad& \mathcal{X}^{i}=\emptyset, \\ r^{j,i}f^{j,i}(\boldsymbol{x}) \quad & \mathcal{X}^{i}=\{\boldsymbol{x}\}, \\ 0 \quad & \mathrm{otherwise},
\end{cases}
\end{equation} 
where $r^{j,i} \in [0,1]$ is the existence probability, and $f^{j,i}(\cdot)$ is the state density. A larger $r^{j,i}$ means that the landmark is more likely to be present. If  $r^{j,i}=0$, the corresponding landmark does not exist, and if  $r^{j,i}=1$, the corresponding landmark surely exists. The \ac{MBM} density can be parameterized as $\{w^{j},\{r^{j,i},f^{j,i}(\boldsymbol{x})\}_{i\in \mathbb{I}^{j}}\}_{j\in \mathbb{J}}$. Following the convolution formula \cite[eq.~(4.17)]{mahler2014advances}, the \ac{PMBM} density $f(\mathcal{X})$ is given by
\begin{equation}
    f(\mathcal{X})=\sum_{\mathcal{X}_{\mathrm{U}}\biguplus\mathcal{X}_{\mathrm{D}}=\mathcal{X}}f_{\mathrm{P}}(\mathcal{X}_{\mathrm{U}})f_{\mathrm{MBM}}(\mathcal{X}_{\mathrm{D}}),\label{PMBM}
\end{equation}
which can be completely parameterized by $\lambda(\boldsymbol{x})$ and $\{w^{j},\{r^{j,i},f^{j,i}(\boldsymbol{x})\}_{i\in \mathbb{I}^{j}}\}_{j\in \mathbb{J}}$.  Please note that if there is only one mixture component in the \ac{MBM}, i.e., there is only one global \ac{DA}, then \eqref{MBM} reduces to an \ac{MB}, and \eqref{PMBM} reduces to a \ac{PMB}. If there are no detected landmarks ($\mathcal{X}_{\mathrm{D}}=\emptyset$), \eqref{PMBM} reduces to a PPP.

\section{Graph PMBM-SLAM Algorithm} \label{Sec:framework}

This section introduces the proposed Graph PMBM-SLAM algorithm, which  combines  \ac{RFS} and GraphSLAM. The framework seeks to leverage the advantages of both methods to obtain a \ac{SLAM} solution, where the \ac{RFS} posterior serves for an elegant and theoretically sound treatment of the \ac{DA} uncertainties, and the GraphSLAM serves as a computationally efficient and robust backend algorithm, conditioned on a \ac{DA} and existences of landmarks.

 \vspace{-2mm}

\subsection{Joint Posterior Expressions} \label{Sec:posterior}

\subsubsection{Sensor Trajectory and Map Posterior}

The posterior distribution is denoted by $f(\mathcal{X}, \boldsymbol{s}_{0:K} \mid \mathcal{Z}_{1:K})$, where $\boldsymbol{s}_{0:K}$ denotes the sensor trajectory, and $\mathcal{Z}_{1:K}=(\mathcal{Z}_{1},\; \ldots, \; \mathcal{Z}_{K})$ denotes the measurement batch (i.e., the sequence of measurements up to time step $K$). We can factorize $f(\mathcal{X}, \boldsymbol{s}_{0:K} \mid \mathcal{Z}_{1:K})$ as 
\begin{align}
 f(\boldsymbol{s}_{0:K},& \mathcal{X}|\mathcal{Z}_{1:K} ) = \label{jointposter_mark}\\ 
&  \frac{ f(\boldsymbol{s}_{0})f(\mathcal{X}) \prod_{k=1}^{K}f(\boldsymbol{s}_{k}|\boldsymbol{s}_{k-1})   g(\mathcal{Z}_{1:K} |\boldsymbol{s}_{1:K},\mathcal{X})}{f(\mathcal{Z}_{1:K})},\nonumber
\end{align}
where $f(\boldsymbol{s}_{0})$ denotes the sensor prior density, $f(\mathcal{X})$ denotes the prior set density of the landmark set, $f(\boldsymbol{s}_{k}|\boldsymbol{s}_{k-1})$ was introduced in \eqref{transition_density}, $g(\mathcal{Z}_{1:K} |\boldsymbol{s}_{1:K},\mathcal{X})$ denotes the likelihood function of  measurement batch $\mathcal{Z}_{1:K}$ given $\boldsymbol{s}_{1:k}$ and $\mathcal{X}$, and $f(\mathcal{Z}_{1:K})$ is the normalizing factor. By assuming that the prior is a PPP \cite{fatemi2017poisson} and plugging all these expressions into \eqref{jointposter_mark}, the joint posterior can be expressed in a more explicit form. 

We first proceed to define the required notation. Let $j\in \mathbb{J}$ denote a global hypothesis, which is a valid partition of the set of measurements across all time steps, with each measurement augmented by its respective time step. This set of measurements can be directly obtained from $\mathcal{Z}_{1:K}$. Then, each cell in the $j$-th partition contains the measurements associated to the same unique origin $\mathcal{Y}^{j,i}$, whose measurement sequence is $\mathcal{Z}_{1:K}^{j,i}=(\mathcal{Z}^{j,i}_{1}, \ldots,\mathcal{Z}^{j,i}_{K})$, where $\mathcal{Z}_{k}^{j,i}$ denotes the  measurement set from the source $\mathcal{Y}^{j,i}$ at time step $k$. As the landmarks can only create one measurement per time step, we have that  $|\mathcal{Z}_{k}^{j,i}| \leq 1$. When $\mathcal{Y}^{j,i}=\emptyset$ and $|\mathcal{Z}_{1:K}^{j,i}| = 1$, $\mathcal{Z}_{1:K}^{j,i}$ contains a single clutter measurement. Overall, it holds that $\mathcal{Z}_{1:K}=( \uplus_{i}\mathcal{Z}^{j,i}_{1}, \ldots,  \uplus_{i}\mathcal{Z}^{j,i}_{K} )$.
\begin{theorem}
    \label{theo:posterior}
The joint posterior \eqref{jointposter_mark} can be expressed as
\begin{align}
&f(\boldsymbol{s}_{0:K},\mathcal{X}|\mathcal{Z}_{1:K} ) = e^{-\int \lambda(\boldsymbol{x})\text{d}\boldsymbol{x}-K\int c(\boldsymbol{z}) \text{d} \boldsymbol{z}} \label{jointposter_mark_new}\\ & \sum_{j\in \mathbb{J}}\sum_{\mathcal{X}_{\mathrm{U}}\biguplus\mathcal{Y}^{j,1}\biguplus\dots\biguplus\mathcal{Y}^{j,|\mathbb{I}^{j}|}=\mathcal{X}}\prod_{\boldsymbol{x} \in \mathcal{X}_{\mathrm{U}}}\left(p_{\text{U}}(\boldsymbol{x},\boldsymbol{s}_{1:K})\lambda(\boldsymbol{x}) \right)f(\boldsymbol{s}_{0})\nonumber\\ &  \prod_{k=1}^{K}f(\boldsymbol{s}_{k}|\boldsymbol{s}_{k-1})\prod_{i=1}^{|\mathbb{I}^{j}|}(t(\mathcal{Z}_{1:K}^{j,i}|\boldsymbol{s}_{1:K},\mathcal{Y}^{j,i})\lambda(\mathcal{Y}^{j,i}))/f(\mathcal{Z}_{1:K}).\nonumber
\end{align}
In \eqref{jointposter_mark_new}, $\mathcal{X}_{\mathrm{D}}=\mathcal{Y}^{j,1}\biguplus\dots\biguplus\mathcal{Y}^{j,|\mathbb{I}^{j}|}$ are all detected landmarks, 
and $p_{\text{U}}(\boldsymbol{x},\boldsymbol{s}_{1:K})=\prod_{k=1}^{K}(1-p_{\text{D}}(\boldsymbol{x},\boldsymbol{s}_{k}))$ denotes the misdetection probability for landmarks that have not been detected for the whole time period. Moreover,  $t(\mathcal{Z}_{1:K}^{j,i}|\boldsymbol{s}_{1:K},\mathcal{Y}^{j,i})$ denotes the  likelihood of $\mathcal{Z}_{1:K}^{j,i}$ and is given by
\begin{align} \label{likelihood_single_tar}
&t(\mathcal{Z}_{1:K}^{j,i}|\boldsymbol{s}_{1:K},\mathcal{Y}^{j,i})= \\ &\quad
\begin{cases}
c(\boldsymbol{z}) \quad & |\mathcal{Z}_{1:K}^{j,i}|=1,\mathcal{Y}^{j,i}=\emptyset, \\ \prod_{k=1}^{K} \ell(\mathcal{Z}_{k}^{j,i} |\boldsymbol{s}_{k},\boldsymbol{x}^i)\quad& |\mathcal{Z}_{1:K}^{j,i}|\geq 1,\mathcal{Y}^{j,i}=\{\boldsymbol{x}^i\}, \\0 \quad & \mathrm{otherwise},
\end{cases}\nonumber
\end{align}
where 
\begin{align} \label{likelihood_single_mea}
    \ell(\mathcal{Z}_{k}^{j,i} |\boldsymbol{s}_{k},\boldsymbol{x}^i)=
\begin{cases}
 1-p_{\text{D}}(\boldsymbol{x}^{i},\boldsymbol{s}_{k}) \quad & \mathcal{Z}_{k}^{j,i}=\emptyset, \\ p_{\text{D}}(\boldsymbol{x}^{i},\boldsymbol{s}_{k})f(\boldsymbol{z}|\boldsymbol{x}^{i}\boldsymbol{s}_{k}) \quad& \mathcal{Z}_{k}^{j,i}=\{\boldsymbol{z} \}, \\0 \quad & \mathrm{otherwise}.
\end{cases}
\end{align}
Finally, $\lambda(\mathcal{Y}^{j,i})$ in \eqref{jointposter_mark_new} denotes the prior intensity defined on the set $\mathcal{Y}^{j,i}$ given by
\begin{align} \label{intensity_set}
    \lambda(\mathcal{Y}^{j,i})=
\begin{cases}
 1 \quad & \mathcal{Y}^{j,i}=\emptyset, \\\lambda(\boldsymbol{x}^i) \quad& \mathcal{Y}^{j,i}=\{\boldsymbol{x}^i \}, \\0 \quad & \mathrm{otherwise}.
\end{cases}
\end{align}

\end{theorem}
\begin{proof}
    See Appendix~\ref{prove_PMBM}.
\end{proof}

For using GraphSLAM, a weakness of the introduced notation is that the partition of $\mathcal{Z}_{1:K}$ into $\mathcal{Z}_{1:K}^{j,1}, \ldots , \mathcal{Z}_{1:K}^{j,|\mathbb{I}^{j}|}$ may contain subsets that only comprise a clutter measurement without any corresponding landmarks (see the first entry in \eqref{likelihood_single_tar}), which creates ambiguity in determining the actual number of landmarks. To address this, 
we introduce an auxiliary variable $ \psi^{j,i}\in \{ 0,1\}$, where\footnote{The usage of $\psi^{j,i}$ is similar to the expansion of a Bernoulli density into the sum of two Bernoulli densities with deterministic target existences in the \ac{MBM}01 representation in \cite[Section IV.A]{garcia2018poisson}.} $ \psi^{j,i}=1$ indicates that $\mathcal{Y}^{j,i}$ is non-empty so that the corresponding landmark exists, while $\psi^{j,i}=0$ indicates that $\mathcal{Y}^{j,i}$ is empty so that the corresponding landmark does not exist. Finally, $p(\psi^{j,i}|\boldsymbol{s}_{1:K})$ is the probability that the landmark either exists or not, without depending on any measurements. With this auxiliary variable, 
we can write \eqref{likelihood_single_tar} and  \eqref{intensity_set} as an \ac{MBM}
\begin{align}
t(\mathcal{Z}_{1:K}^{j,i}& |\boldsymbol{s}_{1:K},\mathcal{Y}^{j,i})\lambda(\mathcal{Y}^{j,i})=\label{singleBern_to_multi}\\
&p(\psi^{j,i}=0|\boldsymbol{s}_{1:K})\tilde{f}(\mathcal{Z}_{1:K}^{j,i},\mathcal{Y}^{j,i}|\boldsymbol{s}_{1:K},\psi^{j,i}=0)\nonumber\\   +& p(\psi^{j,i}=1|\boldsymbol{s}_{1:K})\tilde{f}(\mathcal{Z}_{1:K}^{j,i},\mathcal{Y}^{j,i}|\boldsymbol{s}_{1:K},\psi^{j,i}=1),\nonumber
\end{align}
where
\begin{align} \label{joint_mea_land}
    \tilde{f}(&\mathcal{Z}_{1:K}^{j,i},\mathcal{Y}^{j,i}|\boldsymbol{s}_{1:K},\psi^{j,i})=\\&
\begin{cases}
 (1-\psi^{j,i})c(\boldsymbol{z}) \quad & \mathcal{Y}^{j,i}=\emptyset, \\ \psi^{j,i}\prod_{k=1}^{K} \ell(\mathcal{Z}_{k}^{j,i} |\boldsymbol{s}_{k},\boldsymbol{x}^i)\lambda(\boldsymbol{x}^i) \quad& \mathcal{Y}^{j,i}=\{\boldsymbol{x}^i \}, \\0 \quad & \mathrm{otherwise}.
\end{cases}\nonumber
\end{align}
Then, we have
\begin{align}
&f(\boldsymbol{s}_{0:K},\mathcal{X}|\mathcal{Z}_{1:K} ) = \frac{e^{-\int \lambda(\boldsymbol{x})\text{d}\boldsymbol{x}-K\int c(\boldsymbol{z}) \mathrm{d} \boldsymbol{z}}}{f(\mathcal{Z}_{1:K})}   \label{jointposter_mark_new2}\\ & \sum_{j\in \mathbb{J}}\sum_{\mathcal{X}_{\mathrm{U}}\biguplus\mathcal{Y}^{j,1}\biguplus\dots\biguplus\mathcal{Y}^{j,|\mathbb{I}^{j}|}=\mathcal{X}}\prod_{\boldsymbol{x} \in \mathcal{X}_{\mathrm{U}}}\left(p_{\text{U}}(\boldsymbol{x},\boldsymbol{s}_{1:K})\lambda(\boldsymbol{x})\right) f(\boldsymbol{s}_{0})\nonumber\\ &  \prod_{k=1}^{K}f(\boldsymbol{s}_{k}|\boldsymbol{s}_{k-1})\prod_{i=1}^{|\mathbb{I}^{j}|}\sum_{\psi^{j,i}}p(\psi^{j,i}|\boldsymbol{s}_{1:K})\tilde{f}(\mathcal{Z}_{1:K}^{j,i},\mathcal{Y}^{j,i}|\boldsymbol{s}_{1:K},\psi^{j,i}).\nonumber
\end{align}

\subsubsection{Sensor Trajectory, Map, and \ac{DA} Posterior}

We now proceed to express the joint posterior of the map, the sensor trajectory, and the \ac{DA}. We introduce $\mathcal{A}$ as the partition of $\mathcal{Z}_{1:K}$ into $\mathcal{Z}_{1:K}^{j,1}, \ldots , \mathcal{Z}_{1:K}^{j,|\mathbb{I}^{j}|}$, which corresponds to a valid \ac{DA}, and  $\boldsymbol{\psi}^{j}=[\psi^{j,1},\dots,\psi^{j,|\mathbb{I}^{j}|}]$ with $\psi^{j,i}\in \{0,1\}, \forall i\in\{1,\dots,|\mathbb{I}^{j}|\}$, which describes the existence of each corresponding landmark of $\mathcal{Z}_{1:K}^{j,1}, \ldots , \mathcal{Z}_{1:K}^{j,|\mathbb{I}^{j}|}$. By further introducing $\tilde{\mathcal{A}}=(\mathcal{A},\boldsymbol{\psi}^{j})$ as an auxiliary variable in \eqref{jointposter_mark_new2}, we have
\begin{align}
&f(\boldsymbol{s}_{0:K},\mathcal{X},\tilde{\mathcal{A}}|\mathcal{Z}_{1:K} ) =\frac{e^{-\int \lambda(\boldsymbol{x})\text{d}\boldsymbol{x}-K\int c(\boldsymbol{z}) \mathrm{d} \boldsymbol{z}}}{f(\mathcal{Z}_{1:K})}  \label{jointposter_mark_new3}\\ & \sum_{\mathcal{X}_{\mathrm{U}}\biguplus\mathcal{Y}^{j,1}\biguplus\dots\biguplus\mathcal{Y}^{j,|\mathbb{I}^{j}|}=\mathcal{X}}\prod_{\boldsymbol{x} \in \mathcal{X}_{\mathrm{U}}}\left(p_{\text{U}}(\boldsymbol{x},\boldsymbol{s}_{1:K})\lambda(\boldsymbol{x})\right)f(\boldsymbol{s}_{0})\nonumber\\ &  \prod_{k=1}^{K}f(\boldsymbol{s}_{k}|\boldsymbol{s}_{k-1})\prod_{i=1}^{|\mathbb{I}^{j}|}p(\psi^{j,i}|\boldsymbol{s}_{1:K})\tilde{f}(\mathcal{Z}_{1:K}^{j,i},\mathcal{Y}^{j,i}|\boldsymbol{s}_{1:K},\psi^{j,i}).\nonumber
\end{align}

\begin{figure*}
    \centering
\includegraphics[width=0.80\linewidth]{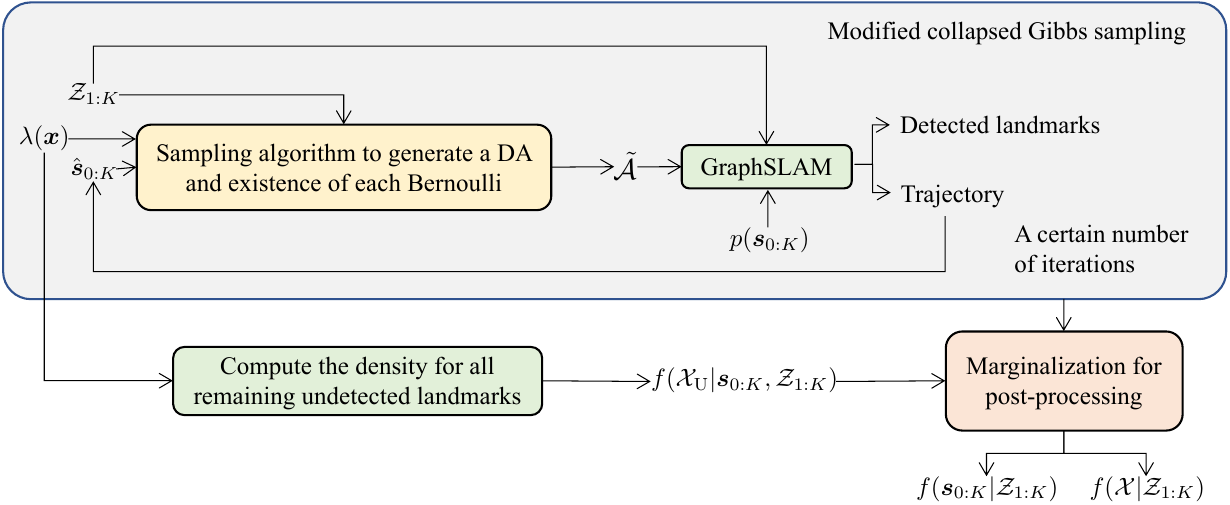}
    \caption{The flowchart of the proposed Graph PMBM-SLAM algorithm. We generate a \ac{DA} using a sampling algorithm, which is conditioned on the sensor trajectory from the last iteration (or the prior trajectory). Then we sample the existence probabilities of all the resulting Bernoulli components. Conditioned on the sampled \ac{DA} and the existence of each Bernoulli, we apply the GraphSLAM algorithm to estimate the sensor trajectory and the detected landmarks. We iterate these two steps for a certain number of times. Finally, we merge  GraphSLAM results from the last $\Gamma$ iterations to output the estimates of the sensor trajectory and the map.  
    }
    \label{fig:flow_batch}
\end{figure*}

\subsection{Overall Framework}
To determine the posterior density $f(\boldsymbol{s}_{0:K},\mathcal{X}|\mathcal{Z}_{1:K} )$, we take inspiration from the collapsed Gibbs sampling technique \cite{robert1999monte,liu1994collapsed}. The core idea of the paper is to iteratively update 1) the DAs, and 2) the map, and the sensor trajectory. In principle, these two steps can be executed either through sampling, as in a Gibbs sampling algorithm, or through optimization, as in a coordinate descent algorithm. In our proposed approach, we sample the DAs and optimize the map and sensor trajectory using GraphSLAM. However, other combinations of these steps are also possible. We refer to the method as a modified collapsed Gibbs sampling algorithm. The term ``collapsed" indicates that we condition the sampling step only on the sensor trajectory, with the map analytically marginalized, and ``modified" denotes that we estimate the sensor trajectory instead of sampling it. The modified collapsed Gibbs sampling iterates the following two stages:

\begin{enumerate}
    \item \emph{Sampling \acp{DA} (See Section \ref{data_associatio_section}):} Sample a candidate $\tilde{\mathcal{A}}$ value from  $f(\tilde{\mathcal{A}}|\boldsymbol{s}_{0:K},\mathcal{Z}_{1:K})$ based on the latest estimate of $\boldsymbol{s}_{0:K}$. 
    \item \emph{GraphSLAM (See Section \ref{Sec:GraphSLAM_givenDA}):} Perform the GraphSLAM algorithm on $f(\boldsymbol{s}_{0:K},\mathcal{X}_{\text{D}}|\mathcal{Z}_{1:K},\tilde{\mathcal{A}})$ to obtain conditional posteriors of the detected landmarks and a sensor trajectory, for the sampled $\tilde{\mathcal{A}}$;
\end{enumerate}

The final sensor trajectory and the map are acquired through a post-processing step, which involves merging the SLAM results from different iterations and considering the undetected landmarks. The corresponding flowchart of the Graph PMBM-SLAM algorithm is summarized in Fig. \ref{fig:flow_batch}.

\section{Data Association Sampling}\label{data_associatio_section}
In this section, the batch \ac{DA} sampling problem is described. First, the DA representation is introduced, and then its weight is computed. 
To simplify the sampling process, instead of directly sampling $\tilde{\mathcal{A}}$, we firstly sample $\mathcal{A}$ from $f(\mathcal{A}|\mathcal{Z}_{1:K},\boldsymbol{s}_{0:K})$, and then sample $\boldsymbol{\psi}$ from $f(\boldsymbol{\psi}|\mathcal{Z}_{1:K},\boldsymbol{s}_{0:K},\mathcal{A})$.

\subsection{Data Association Representation}\label{data_association}
Each \ac{DA}  is a valid assignment of the measurements to their sources (landmarks or clutter), which is equivalent to partition $\mathcal{Z}_{1:K}$ into valid non-empty subsets according to sources, i.e., $\mathcal{Z}^{j,1}_{1:K}, \ldots,\mathcal{Z}^{j,|\mathbb{I}^{j}|}_{1:K}$ for the $j$-th \ac{DA} in \eqref{jointposter_mark_new}. In this section, we  index the measurements in  $\mathcal{Z} _{1:K}$ by $\boldsymbol{m} \in \mathbb{M}$, where $\boldsymbol{m} = (k,\alpha_{k})$, with $k \in \{1,\dots,K\}$ representing the time index and $\alpha_{k} \in \{1,\dots,|\mathcal{Z} _{k}|\}$ representing the index of a measurement in scan $k\leq K$ \cite{williams2015marginal}.  
A DA can now be equivalently viewed as a valid partition of $\mathbb{M}$ into nonempty disjoint index subsets. Each subset (called a \emph{cell} in this paper) contains indices of all measurements from the same source. Hence,  consider $ \mathcal{Z}^{j,i}_{1:K}$, then the $i$-th cell of global hypothesis $j$ is $\mathcal{C}^{j,i}=\{\boldsymbol{m}| \boldsymbol{z}_{\boldsymbol{m}} \in \mathcal{Z}_{1:K}^{j,i} \}$.

A valid \ac{DA} must satisfy several criteria: (i) each measurement can be associated with at most one landmark, so that two cells should be disjoint; (ii) due to the point object assumption, a landmark cannot generate more than one measurement at each time step. Therefore, any cell cannot contain more than one measurement index with the same time index; (iii)  the union of all cells is the index space $\mathbb{M}$. In summary, a valid partition $\mathcal{A}^{j} \in \mathbb{A} $ should satisfy 
\begin{align}
\nonumber  \mathcal{A}^{j}& =\{\{\mathcal{C}^{j,1},\dots,\mathcal{C}^{j,|\mathbb{I}^{j}|}\} 
|\\
& |\bigcup_{\boldsymbol{m}\in \mathcal{C}^{j,\beta}} \{ \boldsymbol{m}| [\boldsymbol{m}]_1=k\}| \leq 1 , \forall \beta, \forall k; \label{valid_partition}\\
&\mathcal{C}^{j,\beta}\cap \mathcal{C}^{j,\gamma} = \emptyset, \forall \beta \neq \gamma; \bigcup_{\beta}\mathcal{C}^{j,\beta}=\mathbb{M}\}, \nonumber
\end{align}
where $\beta, \gamma \in \{1,\dots,|\mathbb{I}^{j}|\}$.
\begin{example}
Let $\mathbb{M}=\{(1,1),(1,2),(2,1),(2,2),(3,1),(3,2)\}$.  One possible \ac{DA} could be $\{\{(1,1),(2,1),(3,2)\}, \{(2,2),(3,1)\}, \{(1,2)\}\}$, so that a landmark is detected at all time steps 1, 2 and 3 with measurements $\boldsymbol{z}_{(1,1)}$, $\boldsymbol{z}_{(2,1)}$ and $\boldsymbol{z}_{(3,2)}$, respectively; another landmark is misdetected at time step 1 and then detected with measurements $\boldsymbol{z}_{(2,2)}$ and $\boldsymbol{z}_{(3,1)}$ at time step 2 and 3, respectively; measurement $\boldsymbol{z}_{(1,2)}$ can  either be a clutter or from a different landmarks. The partition is permutation invariant, so that different orders of cells or different orders of elements in cells do not create a new partition. 
\end{example}

\subsection{Data Association Weight}\label{DA_weight}
Here, we compute the DA hypothesis weight. The measurement set sequence $\mathcal{Z}^{j,i}_{1:K}= (\mathcal{Z}^{j,i}_{1}, \ldots,\mathcal{Z}^{j,i}_{K})$ of the cell $\mathcal{C}^{j,i}$ contains all measurements associated to the same source over time, with $\mathcal{Z}^{j,i}_{k}=\{ \boldsymbol{z}_{\boldsymbol{m}}|[\boldsymbol{m}]_1=k, \boldsymbol{m} \in \mathcal{C}^{j,i}\}$. 
Therefore, once $\{\mathcal{C}^{j,1},\dots,\mathcal{C}^{j,|\mathbb{I}^{j}|}\}$ is determined, the split of the measurement batch $\mathcal{Z}_{1:K}=( \mathcal{Z}^{j,1}_{1:K}, \ldots,  \mathcal{Z}_{1:K}^{j,|\mathbb{I}^{j}|} )$ is determined, and vice versa. The weight  $f(\mathcal{A}^{j}|\mathcal{Z}_{1:K},\boldsymbol{s}_{1:K})$,  is equivalent to $f(\{\mathcal{C}^{j,1},\dots,\mathcal{C}^{j,|\mathbb{I}^{j}|}\}|\boldsymbol{s}_{1:K},\mathcal{Z}_{1:K})$, which we will denote as $w^{j}$ for notational convenience, and is given by
\begin{align}
w^{j}=f(\mathcal{A}^{j}|\boldsymbol{s}_{1:K},\mathcal{Z}_{1:K})\propto\prod_{i=1}^{|\mathbb{I}^{j}|}l^{j,i}\label{weights_section3},
\end{align}
where the proportionality constant, given by the normalizing constant of the factor $f(\mathcal{Z}_{1:K})$ in \eqref{jointposter_mark_new}, ensures that 
$\sum_{j\in\mathbb{J}}w^{j}=1$, and $l^{j,i}=f(\mathcal{Z}_{1:K}^{j,i}|\boldsymbol{s}_{1:K})$, which can be obtained by applying set integral \cite[eq.~(4)]{williams2015marginal}  on \eqref{likelihood_single_tar} over $\mathcal{Y}^{j,i}$, as
\begin{align}
f(\mathcal{Z}_{1:K}^{j,i}|\boldsymbol{s}_{1:K})&= \int t(\mathcal{Z}_{1:K}^{j,i}|\boldsymbol{s}_{1:K},\mathcal{Y}^{j,i})\lambda(\mathcal{\mathcal{Y}}^{j,i})\delta \mathcal{\mathcal{Y}}^{j,i}
\label{compute_deno}\\ &=
\begin{cases}
c(\boldsymbol{z})+ \langle \prod_{k=1}^{K} \ell_{k}^{j,i};\lambda \rangle \quad& |\mathcal{Z}_{1:K}^{j,i}|=1,\\ \langle \prod_{k=1}^{K} \ell_{k}^{j,i};\lambda \rangle \quad & |\mathcal{Z}_{1:K}^{j,i}|>1, \\ 0 \quad & \mathrm{otherwise},
\end{cases}\nonumber
\end{align}
where $\langle \prod_{k=1}^{K} \ell_{k}^{j,i};\lambda \rangle=\langle \prod_{k=1}^{K} \ell(\mathcal{Z}_{k}^{j,i} |\boldsymbol{s}_{k},\boldsymbol{x}^i);\lambda(\boldsymbol{x}^i) \rangle$.

\subsection{Data Association Sampling} \label{Sec:sampling}
We will now discuss a method for obtaining a \ac{DA} sample from \eqref{weights_section3}. Due to the intractably large number of possible \acp{DA}, especially for the batch problem, direct sampling from \eqref{weights_section3} is unpractical. We will utilize the Gibbs sampling and the \ac{MH} algorithms. The  Gibbs sampling algorithm may perform poorly with an undesired initial DA, if several cells need to be merged to get the correct \ac{DA}, and the \ac{MH} algorithm needs to pass through intermediate \acp{DA} with comparatively lower likelihood before forming larger cells. Therefore, we propose a new batch DA sampling algorithm for a point object model, which combines the Gibbs sampling and the \ac{MH} algorithms, and is summarized in Algorithm \ref{alg:proposed}. Note that the algorithm discards a certain number of iterations due to the burn-in period of MCMC sampling. The proposed algorithm takes the advantages of both the Gibbs sampling and the MH algorithms, which can not only handle groups of measurements but also form larger cells before passing through the \ac{MH} algorithm. Both algorithms are described in detail next. 

\begin{algorithm}[t!]
\caption{DA Sampling} \label{alg:proposed}
\begin{algorithmic}[1]
\Require \parbox[t]{\dimexpr\linewidth- \algorithmicindent * 1}{Measurement $\mathcal{Z}_{1:K}$, index set $\mathbb{M}$, sensor trajectory $\boldsymbol{s}_{1:K}$, initial \ac{DA} $\mathcal{A}^{'}$;
\strut}
\Ensure \ac{DA} $\mathcal{A}^{*}$;
\Repeat
\State Gibbs sampling (Algorithm \ref{alg:Gibbs});
\State MH sampling (Algorithm \ref{alg:MH});
\Until{A certain number of iterations;}
\State Return the last sample as  $\mathcal{A}^{*}$.
\end{algorithmic}
\end{algorithm}

\subsubsection{Gibbs Sampling}

We denote the sample at the $\iota$-th iteration of the Gibbs sampler as $\mathcal{A}^{\iota}$, which is a valid partition sampled from the previous sample or input as the initialization. The cells in $\mathcal{A}^{\iota}$ can be indexed by $\{1,\dots,|\mathcal{A}^{\iota}|\}$. To obtain the $(\iota+1)$-th samples using Gibbs sampler from $\mathcal{A}^{\iota}$, we firstly take a single measurement index $\boldsymbol{m} \in \mathcal{C}^{\iota,\beta} \in \mathcal{A}^{\iota}$, from cell  $\mathcal{C}^{\iota,\beta}$. Then, we consider all possible moves/actions of the index (the case of no move is included in the two actions):
\begin{itemize}
    \item Swap $\boldsymbol{m}$ from $\mathcal{C}^{\iota,\beta}$ with the index which has the same time index as $\boldsymbol{m}$ in the $\gamma$-th cell, $\gamma \in \{1,\dots,|\mathcal{A}^{\iota}|\}$, and if no such index exists, this action becomes a move of $\boldsymbol{m}$ from $\mathcal{C}^{\iota,\beta}$ to the $\gamma$-th cell. We denote the new resulting partition as $\mathcal{A}^{\iota}_{\beta\to\gamma}(\boldsymbol{m})$.\footnote{Strictly speaking, it takes the idea of blocked Gibbs sampling \cite{xia2023efficient}, as two items can be changed simultaneously.}
    \item Move $\boldsymbol{m}$ from $\mathcal{C}^{\iota,\beta}$ to a new cell, which was an empty cell before the move, and we denote the new resulting partition as $\mathcal{A}^{\iota}_{\beta\to0}(\boldsymbol{m})$.
\end{itemize}
For notational brevity, we use the shorthand notation $w^{\iota}_{\beta\to\gamma}(\boldsymbol{m})$ to denote the transition probability of forming the new partition $\mathcal{A}^{\iota}_{\beta\to\gamma}(\boldsymbol{m})$. For each of all the possible options, $w^{\iota}_{\beta\to\gamma}(\boldsymbol{m})$ is computed, where a move is sampled from the resulting \ac{PMF} to form the new partition, denoted as 
\begin{align}
&f(\mathcal{A}^{\iota+1}=\mathcal{A}^{\iota}_{\beta\to\gamma}(\boldsymbol{m}) \vert \mathcal{A}^{\iota}, \mathcal{Z}_{1:K},\boldsymbol{s}_{1:K})=w^{\iota}_{\beta\to\gamma}(\boldsymbol{m}),\label{pro_moving}
\end{align}
for $\gamma=0,1,\dots,|\mathcal{A}^{\iota}|$, and $\sum_{\gamma=0}^{|\mathcal{A}^{\iota}|}w^{\iota}_{\beta\to\gamma}(\boldsymbol{m})=1$.

As only two cells in  $\mathcal{A}^{\iota}$ are changed at each sampling time, most of the factors in  \eqref{weights_section3} are common, which reduces the computational cost significantly. In particular, $w^{\iota}_{\beta\to\gamma}(\boldsymbol{m})$ can be calculated more efficiently as
\begin{align}
w^{\iota}_{\beta\to\gamma}(\boldsymbol{m})&\propto \frac{\prod_{i=1}^{|\mathcal{A}^{\iota}|}l^{\iota,i}}{l^{\iota,\beta}l^{\iota,\gamma}}l^{\iota,\beta'}l^{\iota,\gamma'}\propto \frac{l^{\iota,\beta'}l^{\iota,\gamma'}}{l^{\iota,\beta}l^{\iota,\gamma}},\label{weight_efficiently}
\end{align}
where $l^{\iota,\beta}$ and $l^{\iota,\gamma}$ are the likelihood for  $\mathcal{C}^{\iota,\beta}$ and $\mathcal{C}^{\iota,\gamma}$, respectively, and  $l^{\iota,\beta'}$ and $l^{\iota,\gamma'}$ are the likelihood for $\mathcal{C}^{\iota,\beta'}$ and $\mathcal{C}^{\iota,\gamma'}$, respectively, which are the resulting cells after applying the action to $\mathcal{C}^{\iota,\beta}$ and $\mathcal{C}^{\iota,\gamma}$. Also, we have the special cases if $\mathcal{C}^{\iota,\gamma}=\emptyset$, $l^{\iota,\gamma}=1$,  and if $\mathcal{C}^{\iota,\beta'}=\emptyset$, $l^{\iota,\beta'}=1$.\footnote{Some implementation aspects: In \eqref{pro_moving}, there could be some cases resulting $\mathcal{A}^{\iota+1}=\mathcal{A}^{\iota}$, i.e., if $\beta=\gamma$ in general, meaning the selected index stays in the same cell, or if $|\mathcal{C}^{\iota,\beta}|=1$, the swapped cell only has one measurement index that has the same time index as $\boldsymbol{m}$ or 
$\gamma=0$. To avoid considering the same move of the selected measurement index multiple times, we set $P(\mathcal{A}^{\iota+1}=\mathcal{A}^{\iota}_{\beta\to\gamma}(\boldsymbol{m}) \vert \mathcal{A}^{\iota}, ,\mathcal{Z}_{1:K},\boldsymbol{s}_{1:K})=0$ 
when $|\mathcal{C}^{\iota,\beta}|=1$, for any $\gamma$ satisfying $\gamma=\beta$ or $|\mathcal{C}^{\iota,\gamma}|=1$ with its only measurement index having the same time index as $\boldsymbol{m}$. Moreover, there could be some undesired moves of the indices, which create 
unlikely \acp{DA}, i.e., with negligible weight, in the denominator of \eqref{pro_moving}. To reduce the number of considered moves, we can remove undesired moves by setting the corresponding weights to 0.} 
The resulting Gibbs sampling algorithm is summarized in Algorithm \ref{alg:Gibbs}.

\begin{example}
Let $\mathcal{A}^{\iota}=\{\{(1,1),(2,1),(3,2)\}, \{(1,2)\}, \allowbreak\{(2,2),(3,1)\}\}$, and $\mathcal{A}^{\iota}_{1\to 3}((1,1))$ denotes the resulting \ac{DA} of moving the measurement index $(1,1)$ from its original cell to the third cell (the cell $\{(2,2),(3,1)\}$), which is $\mathcal{A}^{\iota}_{1\to 3}((1,1))=\{\{(2,1),(3,2)\}, \{(1,2)\}, \{(1,1),(2,2),(3,1)\}\}$, and the transition probability of such move to forming $\mathcal{A}^{\iota}_{1\to 3}((1,1))$ is $w^{\iota}_{1\to 3}((1,1))$. 
\end{example}
\begin{algorithm}[ht]
\caption{Gibbs Sampling (one iteration)} \label{alg:Gibbs}
\begin{algorithmic}[1]
\Require \parbox[t]{\dimexpr\linewidth- \algorithmicindent * 1}{ Batch measurements $\mathcal{Z}_{1:K}$, index set $\mathbb{M}$, sensor trajectory $\boldsymbol{s}_{1:K}$, \ac{DA} $\mathcal{A}^{\text{in}}$;
\strut}
\Ensure \ac{DA} $\mathcal{A}^{\text{out}}$;
\State Set $\mathcal{A}^{\iota=0}$ as $\mathcal{A}^{\text{in}}$, and $\iota=0$
\For{$n=1:|\mathbb{M}|$} 
\State Calculate transition prob. 
 \eqref{pro_moving} for $\gamma \in \{0,1,\dots,|\mathcal{A}^{\iota}|\}$;
\State Draw sample  $\mathcal{A}^{\iota+1}$;
\State $\iota \gets \iota+1$;
\EndFor
\State Output the last sample as $\mathcal{A}^{\text{out}}$.
\end{algorithmic}
\end{algorithm}

\subsubsection{MH Algorithm}
The Gibbs sampler can be slow as it only takes actions on one measurement index each time (or two for swapping); moreover, it is possible that one index always oscillates between these two cells resulting in a lack of diversity of unlikely DAs \cite{jain2004split}. 
To address these problems, the \ac{MH} algorithm can be used. In the \ac{MH} algorithm, instead of considering the action for a specific index at a sampling time, we consider the split of a cell or the merge of two cells. Since splits and merges change assignments for entire cells at each sampling time step, it leads to a faster algorithm and can avoid oscillations of a single index between these two cells. At each sampling time, if a single cell is chosen, the corresponding cell is considered for a cell split into two sub-cells; if two cells are chosen, the two corresponding cells are considered for a cell merge. There is only one way to merge two cells, while there are multiple ways to split a cell into two sub-cells. To avoid multiple possible split actions, we use the k-means++ algorithm \cite{arthur2007k} to split a cell, as suggested in \cite{granstrom2017likelihood}. 

By following \cite[eq.~(33)-(35)]{granstrom2017likelihood} and choosing the proposal density, the acceptance probabilities of the split of cell $\mathcal{C}^{\iota,\beta}$ into two sub-cells $\mathcal{C}^{\iota,\beta,1}_{\text{split}}$ and $\mathcal{C}^{\iota,\beta,2}_{\text{split}}$, and the merge of two cells $\mathcal{C}^{\iota,\beta}$ and $\mathcal{C}^{\iota,\gamma}$ into a cell $\mathcal{C}^{\iota,\beta,\gamma}_{\text{merge}}$ are given by
\begin{align}
    P\{\text{split}\} &= \min [1,\frac{l^{\iota,\beta,1}_{\text{split}}l^{\iota,\beta,2}_{\text{split}}}{l^{\iota,\beta}}],\label{split_prob}\\
    P\{\text{merge}\} &= \min [1,\frac{l^{\iota,\beta,\gamma}_{\text{merge}}}{l^{\iota,\beta}l^{\iota,\gamma}}],\label{merge_prob}
\end{align}
respectively, with $l^{\iota,\beta,1}_{\text{split}}$, $l^{\iota,\beta,2}_{\text{split}}$,  $l^{\iota,\beta,\gamma}_{\text{merge}}$, $l^{\iota,\beta}$, and $l^{\iota,\gamma}$ denoting the likelihood for the corresponding cells, obtained via \eqref{compute_deno}. Please note that if $|\mathcal{C}^{\iota,\beta}|=1$, the cell cannot be split, resulting in $P\{\text{split}\}=0$ in \eqref{split_prob}; if $\mathcal{C}^{\iota,\beta}$ and $\mathcal{C}^{\iota,\gamma}$ contain indices with the same time index, two cells cannot be merged, resulting in $P\{\text{merge}\}=0$ in \eqref{merge_prob}. 

The interpretation of \eqref{split_prob} and \eqref{merge_prob} is: if the likelihood of the resulting DA is larger than the likelihood of the current DA, the action is for sure performed (with probability 1); if the likelihood of the resulting DA is smaller than the likelihood of the current DA, the action is performed with the probability of the value of the ratio of the likelihood of the resulting DA and the likelihood of the current DA, where we sample on this probability to decide if we perform the action or not. The resulting \ac{MH} algorithm is summarized in Algorithm \ref{alg:MH}.

\begin{algorithm}
\caption{Metropolis-Hastings Algorithm (one iteration)} \label{alg:MH}
\begin{algorithmic}[1]
\Require \parbox[t]{\dimexpr\linewidth- \algorithmicindent * 1}{ Batch measurements $\mathcal{Z}_{1:K}$, index set $\mathbb{M}$, sensor trajectory $\boldsymbol{s}_{1:K}$, \ac{DA} $\mathcal{A}^{\text{in}}$;
\strut}
\Ensure \ac{DA} $\mathcal{A}^{\text{out}}$;
\State Set $\mathcal{A}^{\iota=0}$ as $\mathcal{A}^{\text{in}}$, and $\iota=0$
\For{$n=1:|\mathbb{M}|$} 
\For{$n'=1:|\mathbb{M}|,n'\neq n$}
\If{the $n$-th and $n'$-th indices belong to the same cell in $\mathcal{A}^{\iota}$}
\State Split the cell into two sub-cells; 
\State Compute $P\{\text{split}\}$ 
and 
draw $\varsigma \sim U(0,1)$;
\If{$\varsigma \leq P\{\text{split}\}$}
Split the cell;
\EndIf
\Else
\State Compute $P\{\text{merge}\}$  and 
draw $\varsigma \sim U(0,1)$;
\If{$\varsigma \leq P\{\text{Merge}\}$}
Merge two cells;
\EndIf
\EndIf
\State Set the resulting DA as $\mathcal{A}^{\iota+1}$;
\State $\iota \gets \iota+1$;
\EndFor
\EndFor
\State Output the last sample as $\mathcal{A}^{\text{out}}$.
\end{algorithmic}
\end{algorithm}

\subsection{Sampling  Existence Probabilities} \label{sampling_on_exi}
Given the \ac{DA} $\mathcal{A}^{t}=\{\mathcal{C}^{t,1},\dots,\mathcal{C}^{t,|\mathcal{A}^{t}|}\}$, there are $\mathcal{Y}^{t,1},\dots,\mathcal{Y}^{t,|\mathcal{A}^{t}|}$ Bernoullis in total, as each cell in $\mathcal{A}^{t}$ refers to a Bernoulli. Since $\boldsymbol{\psi}^{t,i}$ indicates whether the corresponding landmark exists or not (the Bernoulli $\mathcal{Y}^{t,i}$ is empty or contains the landmark), its corresponding probabilities are $p(\psi^{t,i}=1\mid \mathcal{A}^{t},\mathcal{Z}_{1:K},\boldsymbol{s}_{0:K})=r^{t,i}_{K}$ and $p(\psi^{t,i}=0\mid \mathcal{A}^{t},\mathcal{Z}_{1:K},\boldsymbol{s}_{0:K})=1-r^{t,i}_{K}$, respectively, with $r^{t,i}_{K}$ indicating the existence probability of $\mathcal{Y}^{t,i}$, given by \cite[eq.~(32)]{fatemi2017poisson}
\begin{align}
r^{t,i}_{K} &=
\begin{cases}
\frac{\langle \prod_{k=1}^{K} \ell_{k}^{t,i};\lambda \rangle}{c(\boldsymbol{z})+ \langle \prod_{k=1}^{K} \ell_{k}^{t,i};\lambda \rangle }\quad& |\mathcal{C}^{t,i}|=1 \quad(|\mathcal{Z}_{1:K}^{t,i}|=1),\\ 1\quad & |\mathcal{C}^{t,i}|>1 \quad(|\mathcal{Z}_{1:K}^{t,i}|>1),\\ 0 \quad & \mathrm{otherwise}.
\end{cases}\label{existence_probability_new}
\end{align}
Considering all components, the \ac{PMF} of $\boldsymbol{\psi}$ is given by 
\begin{align}
f(\boldsymbol{\psi} \mid \mathcal{A}^{t},\mathcal{Z}_{1:K},\boldsymbol{s}_{0:K})= \prod_{i=1}^{|\mathbb{I}^{t}|}p(\psi^{t,i}\mid \mathcal{A}^{t},\mathcal{Z}_{1:K},\boldsymbol{s}_{0:K}).
\end{align}
Therefore, to draw a sample $\boldsymbol{\psi}^{t}$ from $p(\boldsymbol{\psi} \mid \mathcal{A}^{t},\mathcal{Z}_{1:K},\boldsymbol{s}_{0:K})$ is equivalently to sample $\psi^{t,i}$ from $p(\psi^{t,i}\mid \mathcal{A}^{t},\mathcal{Z}_{1:K},\boldsymbol{s}_{0:K})$ for all $i\in\{1,\dots,|\mathcal{A}^{t}|\}$.\footnote{Some implementation aspects: Although there are $|\mathcal{A}^{t}|$ components should be considered and $|\mathcal{A}^{t}|$ is not usually a small number, many of the components are 1, since these landmarks for sure exist and we can directly set the corresponding $\psi^{t,i}$ as 1, which corresponds to the second entry in \eqref{existence_probability_new}. Then, only landmarks that correspond to the first entry in \eqref{existence_probability_new} (the landmark could either be a real landmark or a false alarm caused by clutter) need to be considered, and the number of which is usually not large. To further simplify the problem, we can also only directly set $\psi^{t,i}$ as 0 if the corresponding existence probability is lower than a threshold.}

\section{GraphSLAM Given a Data Association} \label{Sec:GraphSLAM_givenDA}
In Section \ref{data_associatio_section}, we generated a DA sample  $\tilde{\mathcal{A}}$ based on the proposed sampling algorithm and performed sampling on the existence probabilities. In this section, we will focus on how to estimate $\boldsymbol{s}_{0:K}$ and $\mathcal{X}$ from $f(\boldsymbol{s}_{0:K},\mathcal{X}|\mathcal{Z}_{1:K},\tilde{\mathcal{A}})$ with the help of an \ac{MAP} estimator. 

\subsection{Representation}
By fixing the DA and the existence of each Bernoulli in \eqref{jointposter_mark_new}, i.e, conditioning on $\tilde{\mathcal{A}}$, $f(\boldsymbol{s}_{0:K},\mathcal{X}|\mathcal{Z}_{1:K},\tilde{\mathcal{A}})$ follows
\begin{align}
&f(\boldsymbol{s}_{0:K},\mathcal{X}|\mathcal{Z}_{1:K},\tilde{\mathcal{A}})= \frac{e^{-\int \lambda(\boldsymbol{x})\text{d}\boldsymbol{x}-K\int c(\boldsymbol{z}) \mathrm{d} \boldsymbol{z}}}{f(\mathcal{Z}_{1:K}|\tilde{\mathcal{A}})} \nonumber\\ &\sum_{\mathcal{X}_{\mathrm{U}}\biguplus\mathcal{Y}^{j,1}\biguplus\dots\biguplus\mathcal{Y}^{j,|\mathbb{I}^{j}|}=\mathcal{X}}\prod_{\boldsymbol{x} \in \mathcal{X}_{\mathrm{U}}}\left(p_{\text{U}}(\boldsymbol{x},\boldsymbol{s}_{1:K})\lambda(\boldsymbol{x})\right)f(\boldsymbol{s}_{0})  \nonumber\\ & \prod_{k=1}^{K}f(\boldsymbol{s}_{k}|\boldsymbol{s}_{k-1})\prod_{i\in\mathbb{I}^{j}} \tilde{f}(\mathcal{Z}_{1:K}^{j,i},\mathcal{Y}^{j,i}|\boldsymbol{s}_{1:K},\psi^{j,i}).\label{marg_posterior_conditioned_DA_section4}
\end{align}
Once $\mathcal{A}$ is determined, $|\mathbb{I}^{j}|$ is fixed and the union $\mathcal{Y}^{j,1}\biguplus\dots\biguplus\mathcal{Y}^{j,|\mathbb{I}^{j}|}$ indicates there are $|\mathbb{I}^{j}|$ landmark sets in total, where the emptiness of each landmark set $\mathcal{Y}^{j,i}$ (i.e., the existence of each corresponding landmark) is determined by $\psi^{j,i}$. All the remaining landmarks $\mathcal{X}\backslash \mathcal{X}_{\mathrm{D}}$ are part of $\mathcal{X}_{\mathrm{U}}$. 
In \eqref{marg_posterior_conditioned_DA_section4}, $\mathcal{X}_{\mathrm{U}}$ is dependent on $\boldsymbol{s}_{0:K}$ but independent to $(\mathcal{A},\boldsymbol{\psi})$. Therefore, we can obtain $f(\mathcal{X}_{\mathrm{U}}|\boldsymbol{s}_{0:K},\mathcal{Z}_{1:K})$, which is given by
\begin{align}
f(\mathcal{X}_{\mathrm{U}}& |\boldsymbol{s}_{0:K},\mathcal{Z}_{1:K})=\nonumber\\&  e^{-\int p_{\text{U}}(\boldsymbol{x},\boldsymbol{s}_{1:K})\lambda(\boldsymbol{x})\text{d}\boldsymbol{x}}\prod_{\boldsymbol{x} \in \mathcal{X}_{\mathrm{U}}}\left(p_{\text{U}}(\boldsymbol{x},\boldsymbol{s}_{1:K})\lambda(\boldsymbol{x})\right),\label{marg_posterior_conditioned_DA_section6}
\end{align}
which is a \ac{PPP} density as shown in \eqref{PPP}, with its intensity being $p_{\text{U}}(\boldsymbol{x},\boldsymbol{s}_{1:K})\lambda(\boldsymbol{x})$. 
In addition, by expanding $\tilde{f}(\mathcal{Z}_{1:K}^{j,i},\mathcal{Y}^{j,i}|\boldsymbol{s}_{1:K},\psi^{j,i})$ with \eqref{joint_mea_land}, we have
\begin{align}
&f(\boldsymbol{s}_{0:K},\mathcal{Y}^{j,1},\dots,\mathcal{Y}^{j,|\mathbb{I}^{j}|}|\mathcal{Z}_{1:K},\tilde{\mathcal{A}})\propto f(\boldsymbol{s}_{0})\prod_{k=1}^{K}f(\boldsymbol{s}_{k}|\boldsymbol{s}_{k-1})\nonumber\\& \qquad \qquad \times\prod_{i\in\mathbb{I}^{j}:\psi^{j,i}=1}\prod_{k=1}^{K} \ell(\mathcal{Z}_{k}^{j,i} |\boldsymbol{s}_{k},\boldsymbol{x}^i)\lambda(\boldsymbol{x}^i),\label{marg_posterior_conditioned_DA_section7}
\end{align}
where the proportionality  also corresponds to $\tilde{f}(\mathcal{Z}_{1:K}^{j,i},\mathcal{Y}^{j,i}|\boldsymbol{s}_{1:K},\psi^{j,i}=0)=c(\boldsymbol{z})$. As $\boldsymbol{\psi}$ is fixed, $f(\boldsymbol{s}_{0:K},\mathcal{Y}^{j,1},\dots,\mathcal{Y}^{j,|\mathbb{I}^{j}|}|\mathcal{Z}_{1:K},\tilde{\mathcal{A}})$ becomes an MB01 RFS, i.e., it is an \ac{MB} \ac{RFS} where the existence probabilities of all resulting Bernoullis are either 0 or 1 \cite{garcia2018poisson}. Our goal is to obtain  estimates of $\boldsymbol{s}_{0:K}$ and $\mathcal{X}$ from \eqref{marg_posterior_conditioned_DA_section4}.

\subsection{GraphSLAM Approximations} \label{GraphSLAM}
For notational brevity, we drop the DA index $j$ in the following two subsections. To enable the use of GraphSLAM \cite{thrun2006graph}, we apply several approximations.  
First, we note that the \ac{PPP} part is independent of $(\mathcal{A},\boldsymbol{\psi})$  \eqref{marg_posterior_conditioned_DA_section6} and 
in most cases is not informative regarding the sensor state. 
Hence,  we can first compute the \ac{MAP} estimate of $\boldsymbol{s}_{0:K}$ and $\mathcal{Y}^{1},\dots,\mathcal{Y}^{|\mathcal{A}|}$  with  GraphSLAM on \eqref{marg_posterior_conditioned_DA_section7}, and then update the \ac{PPP} intensity according to \eqref{marg_posterior_conditioned_DA_section6}. 
Second, we drop non-existing landmarks based on $\boldsymbol{\psi}$. To this end, we introduce the number of existing landmarks as $\kappa=\sum_{i=1}^{|\mathcal{A}|}\psi^{i}$, and reorder $\mathcal{Y}^{1},\dots,\mathcal{Y}^{|\mathcal{A}|}$  to keep the first $\kappa$ Bernoullis with $\psi^{i}=1$, and the rest $\kappa+1$ to  $|\mathcal{A}|$ Bernoullis with $\psi^{i}=0$. We also introduce the random variable  $\boldsymbol{q}^{i}_{k}=[\boldsymbol{s}^{\mathsf{T}}_{k}, (\boldsymbol{x}^{i})^{\mathsf{T}}]^{\mathsf{T}}$
comprising the sensor state at time $k$ and the state of the corresponding landmark of $\mathcal{C}^{i}$, and the random variable 
$\boldsymbol{q}=[\boldsymbol{s}^{\mathsf{T}}_{0},\boldsymbol{s}^{\mathsf{T}}_{1},\dots,\boldsymbol{s}^{\mathsf{T}}_{K},(\boldsymbol{x}^{1})^{\mathsf{T}},\dots,(\boldsymbol{x}^{\kappa})^{\mathsf{T}}]^{\mathsf{T}}$. 
Then, $\boldsymbol{q}$ can be estimated by maximizing the posterior 
 \begin{align}
 &\arg \underset{\boldsymbol{q}} {\max} ~ f(\boldsymbol{s}_{0:K},\mathcal{Y}^{1},\dots,\mathcal{Y}^{|\mathcal{A}|}|\mathcal{Z}_{1:K},\tilde{\mathcal{A}})\label{eq:optimizationProblem}\\&= \arg \underset{\boldsymbol{q}} {\max}~ f(\boldsymbol{s}_{0})
 \prod_{k=1}^{K} f(\boldsymbol{s}_{k}|\boldsymbol{s}_{k-1})\prod_{i=1}^{\kappa}\prod_{k=1}^{K}\ell(\mathcal{Z}_{k}^i |\boldsymbol{s}_{k},\boldsymbol{x}^i)\lambda(\boldsymbol{x}^{i}).\nonumber 
\end{align}
Third, instead of using $\lambda(\boldsymbol{x}^{i})$, we estimate a Gaussian distribution of $\boldsymbol{x}^{i}$, denoted as $f(\boldsymbol{x}^{i})$, for those $i\leq \kappa$, by following  \cite[Appendix A.C]{ge2022computationally} with mean $\boldsymbol{u}^{i}$ determined by the first detected measurement (the measurement with the smallest time index in $\mathcal{Z}_{1:K}^i$, and we denote its corresponding index as $\boldsymbol{m}_{\text{fir}}^{i}$) and the corresponding sensor state, and very large covariance $\boldsymbol{C}^{i}$. 
Fourth, we approximate $p_{\text{D}}(\boldsymbol{x}^{i},\boldsymbol{s}_{k})$ in $\ell(\mathcal{Z}_{k}^i |\boldsymbol{s}_{k},\boldsymbol{x}^i)$ from \eqref{likelihood_single_mea}
to be a constant $p_{\text{D}}>0$ in the \ac{FOV} of the sensor   
and  0 outside the \ac{FOV} of the sensor. 
Hence, \eqref{eq:optimizationProblem} can be rewritten by 
 \begin{align}
 &\arg \underset{\boldsymbol{q}} {\max} ~f(\boldsymbol{s}_{0})\prod_{k=1}^{K}f(\boldsymbol{s}_{k}|\boldsymbol{s}_{k-1}) \nonumber\\&\quad \quad \quad \times \prod_{i=1}^{\kappa}f(\boldsymbol{x}^{i})\prod_{k=1}^{K}\prod_{\boldsymbol{z}\in\mathcal{Z}_{k}^i:p_{\text{D}}>0 } f(\boldsymbol{z}|\boldsymbol{s}_{k},\boldsymbol{x}^i).\label{eq:optimizationProblem_rewritten} 
\end{align}

\subsection{GraphSLAM Optimization} \label{estimation}
By plugging \eqref{transition_density} and \eqref{pos_to_channelestimation} into \eqref{eq:optimizationProblem_rewritten}, we can solve  $\arg \underset{\boldsymbol{q}} {\min}  ~ {\mathcal{E}}(\boldsymbol{q})$ for the optimization problem in \eqref{eq:optimizationProblem_rewritten}, 
where 
\begin{align}
  {\mathcal{E}}(&\boldsymbol{q})  = (\boldsymbol{s}_{0}-\boldsymbol{\epsilon}_{0})^{\textsf{T}}\boldsymbol{P}_{0}^{-1}(\boldsymbol{s}_{0}-\boldsymbol{\epsilon}_{0}) +  \label{eq:optimizationProblem_min} \\& \sum_{k=1}^{K}(\boldsymbol{s}_{k}-\boldsymbol{v}(\boldsymbol{\epsilon}_{k-1}))^{\textsf{T}}\boldsymbol{Q}^{-1}(\boldsymbol{s}_{k}-\boldsymbol{v}(\boldsymbol{\epsilon}_{k-1}))+ \nonumber\\& \sum_{i=1}^{\kappa}((\boldsymbol{x}^{i}-\boldsymbol{u}^{i})^{\textsf{T}}(\boldsymbol{C}^{i})^{-1}(\boldsymbol{x}^{i}-\boldsymbol{u}^{i})+\nonumber\\& \sum_{k=1}^{K}\sum_{\boldsymbol{z}\in\mathcal{Z}^{i}_{k}:p_{\text{D}}>0}(\boldsymbol{z}-\boldsymbol{h}(\hat{\boldsymbol{q}}^{i}_{k}))^{\textsf{T}}(\boldsymbol{R}_{k}^{i})^{-1}(\boldsymbol{z}-\boldsymbol{h}(\hat{\boldsymbol{q}}^{i}_{k}))),  \notag 
\end{align}
with $\boldsymbol{\epsilon}_{k}$ and $\boldsymbol{P}_{k}$ denoting the mean and the covariance of $\boldsymbol{s}_{k}$ for $k\in\{0,\cdots,K\}$, respectively. To optimize \eqref{eq:optimizationProblem_min}, we start from an estimate $\hat{\boldsymbol{q}}=[\boldsymbol{\epsilon}^{\mathsf{T}}_{0},\boldsymbol{\epsilon}^{\mathsf{T}}_{1},\dots,\boldsymbol{\epsilon}^{\mathsf{T}}_{K},(\boldsymbol{u}^{1})^{\mathsf{T}},\dots,(\boldsymbol{u}^{\kappa})^{\mathsf{T}}]^{\mathsf{T}}$, and apply gradient descent, as detailed in Appendix \ref{sec:gradient}. Here, $\boldsymbol{\epsilon}^{\mathsf{T}}_{k}=\boldsymbol{v}(\boldsymbol{\epsilon}^{\mathsf{T}}_{k-1})$, for $k\ge 1$. After convergence, we obtain the final estimate $\hat{\boldsymbol{q}}$ and the associated information matrix $\boldsymbol{\Omega}$. 
The mean and covariance of $f(\boldsymbol{s}_{0:K}|\mathcal{Z}_{1:K},\tilde{\mathcal{A}})$ are then given by
\begin{align}\label{update_mean_covariance_trajectory}
    &{\boldsymbol{\epsilon}}_{0:K}=  [ \hat{\boldsymbol{q}}]_{ 1:\nu (K+1)}, \\&
    \boldsymbol{P}_{0:K}= [\boldsymbol{\Omega}^{-1}]_{ 1:\nu (K+1),1:\nu (K+1)}.\label{update_mean_covariance_trajectory2}
\end{align}
where $\nu=\text{dim}(\boldsymbol{s}_{k})$. 
Similarly, the mean and covariance of the map are given by
\begin{align}\label{update_mean_covariance_map}
    &\boldsymbol{u}_{\text{map}}=  [ \hat{\boldsymbol{q}}]_{\nu (K+1)+1:\text{end}}, \\&
    \boldsymbol{C}_{\text{map}}= [ \boldsymbol{\Omega}^{-1}]_{\nu (K+1)+1:\text{end},\nu (K+1)+1:\text{end}},\label{update_mean_covariance_map2}
\end{align}
where $\boldsymbol{C}_{\text{map}}$ is generally a full matrix, as landmarks are correlated to each other, when not conditioned on the sensor state trajectory. 
In addition, the updated mean and the covariance of each landmark can be directly obtained from $\boldsymbol{u}_{\text{map}}$ and  $\boldsymbol{C}_{\text{map}}$ by taking the corresponding parts, denoted as
\begin{align}\label{update_mean_covariance_landmark}
    &\boldsymbol{u}^{i}= [ \boldsymbol{u}_{\text{map}}]_{\mu(i-1)+ 1:\mu i}, \\&
    \boldsymbol{C}^{i}= [ \boldsymbol{C}_{\text{map}}]_{\mu(i-1)+ 1:\mu i,\mu(i-1)+ 1:\mu i},\label{update_mean_covariance_landmark2}
\end{align}
with $\mu=\text{dim}(\boldsymbol{x}^{i})$, and the existence probability $r^{i}=1$ since it exists for sure. For the remaining Bernoullis, the existence probability $r^{i}=0$, and the corresponding $\boldsymbol{u}^{i}$ and $\boldsymbol{C}^{i}$ do not exist. Therefore, we only need to output $\{r^{i}=1,\boldsymbol{u}^{i},\boldsymbol{C}^{i}\}_{i\in \{1,\dots,\kappa\}}$  for the map. It is important to note that $\boldsymbol{\Omega}$ usually has a high dimension, and taking the inverse is computationally costly. There are  computationally efficiently methods to compute \eqref{update_mean_covariance_trajectory}--\eqref{update_mean_covariance_landmark2}, e.g. \cite[Section 5.5]{thrun2006graph}.

\section{Marginalization Over Samples}\label{Sec:fuse_over_samples}
In this section, we describe how the final sensor trajectory, the \ac{MB} of detected landmarks, and the \ac{PPP} intensity of undetected landmarks are computed. 

\subsection{Marginalization} \label{sec:MB}
We keep the SLAM results from the last $\Gamma$ iterations. Based on the collapsed Gibbs sampling theory, all $\Gamma$ samples of $\tilde{\mathcal{A}}$, which we kept after the burn-in period, are equivalent to samples that are directly sampled from  $f(\tilde{\mathcal{A}}|\mathcal{Z}_{1:K})$. For a specific sample $\tilde{\mathcal{A}}^{t}$,
GraphSLAM provides ${\boldsymbol{\epsilon}}_{0:K}^{t}$ and $\boldsymbol{P}_{0:K}^{t}$, and $\{r^{t,i},\boldsymbol{u}^{t,i},\boldsymbol{C}^{t,i}\}_{i\in \{1,\dots,\kappa^{t}\}}$ for $f(\boldsymbol{s}_{0:K}|\mathcal{Z}_{1:K},\tilde{\mathcal{A}}^{t})$ and $f(\mathcal{Y}^{1},\dots,\mathcal{Y}^{\kappa^{t}}|\mathcal{Z}_{1:K},\tilde{\mathcal{A}}^{t})$, respectively. Therefore, the desired posterior approximations can be obtained by marginalizing across all $\tilde{\mathcal{A}}^{t}$ samples, with each sample having the same weight, given by
\begin{align}
    &f(\boldsymbol{s}_{0:K}|\mathcal{Z}_{1:K})\approx \frac{1}{\Gamma}\sum_{t=1}^{\Gamma}f(\boldsymbol{s}_{0:K}^{t}|\mathcal{Z}_{1:K},\tilde{\mathcal{A}}^{t}),\label{marg_sta} \\&f(\mathcal{Y}^{1},\dots,\mathcal{Y}^{|\mathbb{I}|}|\mathcal{Z}_{1:K})\approx \label{marg_map} \frac{1}{\Gamma}\sum_{t=1}^{\Gamma}f(\mathcal{Y}^{1},\dots,\mathcal{Y}^{\kappa^{t}}|\mathcal{Z}_{1:K},\tilde{\mathcal{A}}^{t}).
\end{align}

In terms of \eqref{marg_sta}, the final updated trajectory has mean ${\boldsymbol{\epsilon}}_{0:K}\approx {1}/{\Gamma}\sum_{t=1}^{\Gamma} {\boldsymbol{\epsilon}}_{0:K}^{t}$, and covariance  $ \boldsymbol{P}_{0:K}\approx  \frac{1}{\Gamma}\sum_{t=1}^{\Gamma} (\boldsymbol{P}_{0:K}^{t}+ ({\boldsymbol{\epsilon}}_{0:K}^{t}-{\boldsymbol{\epsilon}}_{0:K})({\boldsymbol{\epsilon}}_{0:K}^{t}-{\boldsymbol{\epsilon}}_{0:K})^{\mathsf{T}})$. 
In terms of  \eqref{marg_map}, each sample of the map follows the \ac{MB} distribution, so that \eqref{marg_map} is an \ac{MBM}. To marginalize the \ac{MBM} over all samples into a single \ac{MB}, several practical aspects must be addressed: (i) the numbering of the landmarks across the samples $\tilde{\mathcal{A}}^{t}$; (ii) computation of the spatial density and existence probability of each MB: (iii) pruning and merging. 

To address the first aspect, we introduce a vector to index all landmarks in the resulting \ac{MB} of each sample, denoted as $\boldsymbol{\sigma}^{t}=[\sigma^{t}(1),\dots,\sigma^{t}(\kappa^{t})]$, defined as 
$\sigma^{t}(i)=\boldsymbol{m}_{\text{fir}}^{t,i}$, 
where we recall that $\boldsymbol{m}_{\text{fir}}^{t,i}$ is the index of the first (earliest) measurement in $\mathcal{C}^{t,i}$.\footnote{This implies that we assume that if $\boldsymbol{m}_{\text{fir}}^{t,i}$ are the same, $\mathcal{C}^{t,i}$ are from the same source. It is possible that two cells in two different $\tilde{\mathcal{A}}$ with different $\boldsymbol{m}_{\text{fir}}^{t,i}$ could be still from the same source, where all the measurements assigned to a landmark are the same, expect the first one. Although these two cells are viewed as from different landmarks, they can be merged in the end, as they are close to each other (see later).}
Different samples may have different $\boldsymbol{\sigma}^{t}$, since the source of each $\mathcal{C}^{t,i}$ may be different, which is $f(\boldsymbol{x}^{i})$ generated using the corresponding measurement of $\boldsymbol{m}_{\text{fir}}^{t,i}$.  To make $\boldsymbol{\sigma}^{t}$ consistent in all \acp{DA}, we pick up all unique $\boldsymbol{m}_{\text{fir}}^{t,i}$, and re-index them with $i \in \{1,\dots,|\mathbb{I}|\}$, where $|\mathbb{I}|$ denotes the number of unique $\boldsymbol{m}_{\text{fir}}^{t,i}$ across all $\Gamma$ samples, which represents all different landmarks over all \acp{DA}. Therefore, $\boldsymbol{\sigma}^{t}$ can be extended and rewritten as a vector $\tilde{\boldsymbol{\sigma}}^{t}$ with length $|\mathbb{I}|$ and components $\tilde{{\sigma}}^{t}(i)\in  \{0,1\},\forall i \in \mathbb{I}= \{1,\dots,|\mathbb{I}|\}$, where $\tilde{{\sigma}}^{t}(i)=1$ means the corresponding landmark exists in the $t$-th sample, and $\tilde{{\sigma}}^{t}(i)=0$ means the corresponding landmark non-exists in the $t$-th sample. We also extend and reorder $\{r^{t,i},\boldsymbol{u}^{t,i},\boldsymbol{C}^{t,i}\}_{i\in \{1,\dots,\kappa^{t}\}}$ into $\{r^{t,i},\boldsymbol{u}^{t,i},\boldsymbol{C}^{t,i}\}_{i\in \mathbb{I}}$ by setting $r^{t,i}=0$, if the corresponding $\tilde{{\sigma}}^{t}(i)=0$. Then, the landmark MB for $i \in \mathbb{I}$ can be set to $r^{i}= {\sum_{t=1}^{\Gamma}\tilde{{\sigma}}^{t}(i)}/{ \Gamma}$ and 
\begin{align}
&\boldsymbol{u}^{i}= \frac{1}{\Gamma r^{i} }\sum_{t \in  \{1,\dots,\Gamma\}:\tilde{{\sigma}}^{t}(i) =1} \boldsymbol{u}^{t,i} 
\\&
    \boldsymbol{C}^{i}= \frac{1}{ \Gamma r^{i}}\sum_{t \in  \{1,\dots,\Gamma\}:\tilde{{\sigma}}^{t}(i) =1} (\boldsymbol{C}^{t,i}+ (\boldsymbol{u}^{t,i}-\boldsymbol{u}^{i})(\boldsymbol{u}^{t,i}-\boldsymbol{u}^{i})^{\mathsf{T}})  
\end{align}
After marginalizing over $\tilde{\mathcal{A}}$, 
an updated \ac{MB} to represent the map of all detected landmarks  $\{r^{i},\boldsymbol{u}^{i},\boldsymbol{C}^{i}\}_{i\in \mathbb{I}}$ is acquired. Finally, we prune Bernoullis with low existence probabilities and merge  Bernoullis which are very close to each other. The proposed method provides an efficient way to approximate the MBM into an MB. More accurate MB approximation methods exist, e.g., by finding the best-fitting MB that minimizes the \ac{KL} divergence \cite{williams2014efficient}.

\subsection{PPP intensity for Undetected Landmarks}\label{sec:ppp_updated}

Apart from the detected landmarks, we have the updated \ac{PPP} for all remaining undetected landmarks,  $f(\mathcal{X}_{\mathrm{U}}|\boldsymbol{s}_{0:K},\mathcal{Z}_{1:K})$ 
in \eqref{marg_posterior_conditioned_DA_section6}. We can also marginalize out the sensor trajectory to acquire $f(\mathcal{X}_{\mathrm{U}}|\mathcal{Z}_{1:K})$,  which results in the updated intensity as
\begin{align}
    \check{\lambda}(\boldsymbol{x})=\int f(\boldsymbol{s}_{0:K}|\mathcal{Z}_{1:K})\prod_{k=1}^{K}(1-p_\text{D}(\boldsymbol{s}_{k},\boldsymbol{x}))\lambda(\boldsymbol{x})\text{d}\boldsymbol{s}_{0:K}. \label{updated_PPP}
\end{align}
Together with the marginalized \ac{MB} computed in Section \ref{sec:MB}, the final map is approximated as a \ac{PMB}.

\section{Results} \label{Sec:results}
In this section, we assess the proposed algorithm in a simulated vehicular setting, conducting a comparison with a benchmark. We outline the simulation environment, detail the performance metrics, and describe the benchmark algorithm before delving into an analysis of SLAM outcomes regarding localization and mapping effectiveness.

\subsection{Simulation Environment}
\begin{figure}
\centerline{\includegraphics[width=1\linewidth]{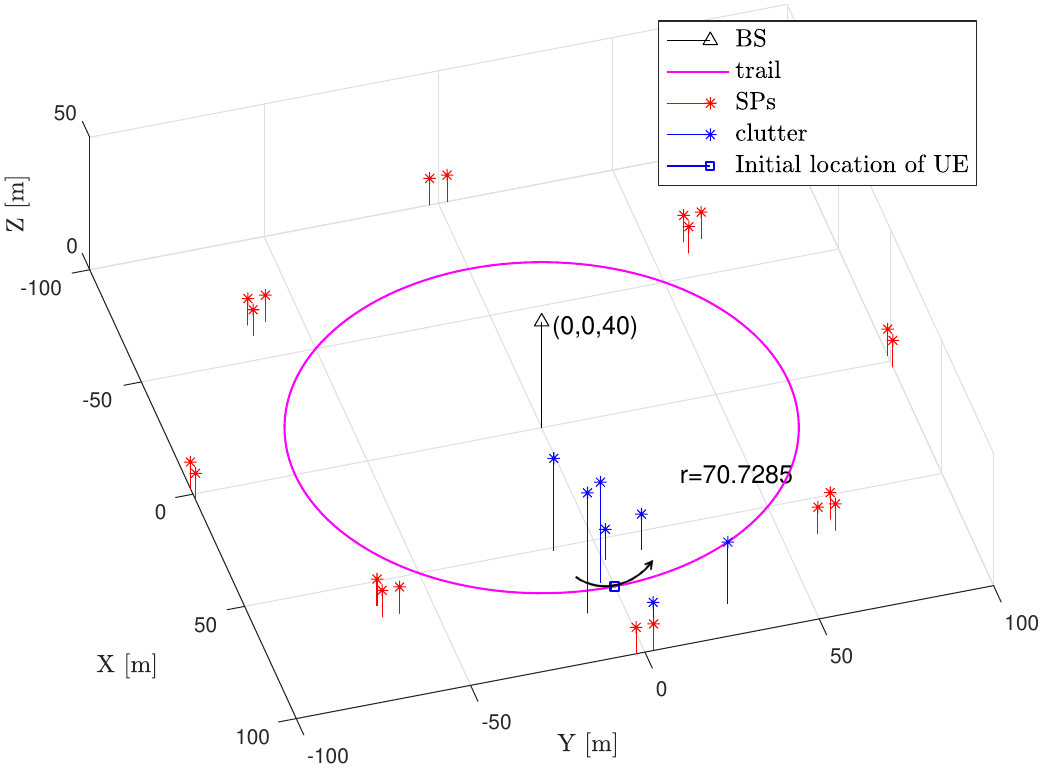}}
\caption{Scenario with the environment of a \ac{bs}, 20 SPs and 7 clutters. The \ac{ue} moves counterclockwise along the trail centered at the \ac{bs}.}
\label{fig:env}
\end{figure}

We consider a propagation environment of bistatic radio SLAM, similar to \cite{ge20205GSLAM,ge2022computationally}, featuring a single vehicle as the \ac{ue}, as shown in Fig~\ref{fig:env}. There is a single \ac{bs} in the environment located at $[0 \, \text{m},0 \, \text{m},40\, \text{m}]^{\mathsf{T}} $, and 20 \acp{sp} in 8 distinct clusters.  
The UE functions as the sensor,  the BS is a known landmark and the \acp{sp} serve as unknown landmarks. The state of the single \ac{ue} $\boldsymbol{s}_{k-1}$ comprises the 3D position  $\boldsymbol{x}_{\mathrm{UE},k-1}=[x_{k-1},y_{k-1},z_{k-1}]^{\mathsf{T}}$, the heading $\varpi_{k-1}$, and clock bias $B_{k-1}$. 
The  \ac{ue} does a counterclockwise constant turn-rate movement around the \ac{bs} on the ground, with $\boldsymbol{v}(\boldsymbol{s}_{k-1})$ in \eqref{transition_density} defined as same as the transition function in \cite[eq.~(63)]{ge2022computationally}. The covariance of the process noise is assumed to be the same for all time steps, denoted as $\boldsymbol{Q}$. The \ac{ue} has a concentrated prior regarding its initial position, but possesses no prior knowledge of the map, except for the BS location and the PPP intensity $\lambda(\boldsymbol{x})=1.5 \times 10^{-5}U_{\text{ENV}}$ for the \acp{sp}, with $U_{\text{ENV}}$ denoting a uniform distribution in the environment. We assume that $p_{\text{D}}=0.9$, where the FOV with respect to the BS is unlimited while for the SPs, it is limited to 50 m around the UE. The measurement function $\boldsymbol{h}(\boldsymbol{x}^{i},\boldsymbol{s}_{k})$  is defined as in \cite[Section~2.2]{ge2024single}, and covariance matrix of the measurement noise is fixed to $\boldsymbol{R}=\text{diag}[ 0.1^{2} \, \text{m}^{2},0.01^{2} \, \text{rad}^{2},0.01^{2} \, \text{rad}^{2},0.01^{2} \, \text{rad}^{2},0.01^{2} \, \text{rad}^{2}]$. The clutter measurement intensity is given by $c(\boldsymbol{z})=\Upsilon U_{\text{FOV}}$, with $U_{\text{FOV}}$ representing a uniform distribution inside the \ac{FOV} and $\Upsilon$ representing the expected number of clutter measurements per time step.

\subsubsection{Scenarios}
Four different scenarios are considered. \emph{Scenario I:} low clutter and low process noise case; \emph{Scenario II:} high clutter and low process noise case, \emph{Scenario III:} low clutter and high process noise case; \emph{Scenario IV:} high clutter and high process noise case. Here,  the low clutter and the high clutter cases stand for cases with $\Upsilon=1$ and  $\Upsilon=5$, respectively, and the low process noise and high process noise cases stand for cases with $\boldsymbol{Q}=\text{diag}([ 0.2^{2} \, \text{m}^{2},0.2^{2} \, \text{m}^{2},0 \, \text{m}^{2},0.001^{2} \, \text{rad}^{2},0.2^{2} \, \text{m}^{2}])$ and $\boldsymbol{Q}=\text{diag}([ 0.2^{2} \, \text{m}^{2},0.2^{2} \, \text{m}^{2},0 \, \text{m}^{2},0.001^{2} \, \text{rad}^{2},0.2^{2} \, \text{m}^{2}]\times8)$, respectively.

\subsubsection{Baselines}
First, we assess the performance of the proposed sampling algorithm, comparing it to the Gibbs sampling algorithm and the \ac{MH} algorithm, in Scenario IV with $\Gamma=100$, which is the most challenging scenario among the four scenarios. Next, we assess the performance of the proposed Graph PMBM-SLAM algorithm with $\Gamma=100$ by conducting a comparative analysis with respect to three baselines: 
  the EK-PMB SLAM filter \cite{ge2022computationally}; 
    the RBP-PHD SLAM filter without optimal importance sampling \cite{kim20205g} using 1000 samples;
    the RBP-PHD SLAM filter with optimal importance sampling \cite{kaltiokallio2022towards} using 1000 samples.
%
\subsubsection{Performance Metrics}
The accuracy of DA is assessed using the average of \acp{NMI} \cite{vinh2009information} between each resulting \ac{DA} and the ground-truth \ac{DA}, where the \ac{NMI} being 1 meaning the resulting \ac{DA} and the ground-truth \ac{DA} contain the same information, i.e, same to each other, and the closer \ac{NMI} is to 1, the more accurate resulting \ac{DA} is. The sensor state estimations are evaluated by the \ac{RMSE} for the UE states over time. The mapping performance is quantified using the generalized optimal subpattern assignment (GOSPA) distance \cite{rahmathullah2017generalized}, where the cut-off distance is set to 5, and the exponent factor is set to 2. In total, we undertake 100 Monte Carlo (MC) simulations for all algorithms, and the final results are obtained by averaging over the independent MC simulations.

\subsection{Results and Discussion}

\subsubsection{DA Accuracy}
We initialize the sample with each measurement forming an individual cell. We measured the proposed sampling algorithm has better performance in accuracy than the Gibbs sampling and the \ac{MH} algorithms, which results in the \ac{NMI} at 0.9971, compared to 0.9679 for Gibbs sampling and 0.9804 for the \ac{MH} algorithm for the Scenario IV. The Gibbs sampling algorithm moves at most two indices at a time, which can cause measurements to oscillate between sub-cells and fail to transfer groups of measurements between cells, especially when each measurement starts as an individual cell, leading to poor DA results. Similarly, the MH algorithm performs poorly with this initialization, since it requires merging several cells to achieve correct DA, but may pass through intermediate DAs with lower likelihoods before forming larger cells. The proposed algorithm combines the Gibbs sampling and the MH algorithms, effectively handling groups of measurements and forming larger cells before using the MH algorithm. The inaccurate DAs result in poor SLAM results, for example, the resulting GOSPA distances are 10.37~m and 7.36~m if the proposed SLAM framework uses only the Gibbs sampling algorithm or the \ac{MH} algorithm, respectively, compared to 1.55~m when the proposed sampling algorithm is used. While this proposed method outperforms the individual algorithms, it still does not perfectly solve the DA problem, as evidenced by its NMI being below 1. The primary reasons for this shortfall are the presence of cluttered scenarios and the misclassification of low-quality measurements as clutters.

\subsubsection{Localization Performance}
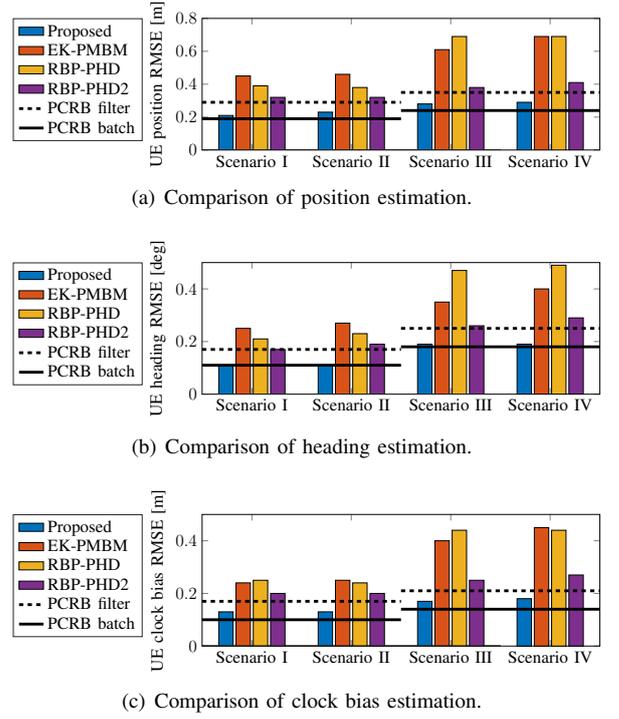
\begin{figure}
     \centering
     \subfigure[Comparison of position estimation.]{         \input{compar_bar_pos}
         \label{Fig.state_compar_pos}}
     \\
     \subfigure[Comparison of heading estimation.]{
     \input{compar_bar_heading}
         \label{Fig.state_compar_heading}
     }
     \\
     \subfigure[Comparison of clock bias estimation.]{
     \input{compar_bar_bias}
         \label{Fig.state_compar_bias}
     }
     \caption{Comparison of sensor trajectory estimation for 4 algorithms under 4 scenarios.}
     \label{Fig.state_compar}
 \end{figure}

Next, the performance of the proposed framework in sensor state estimation is evaluated. Fig.~\ref{Fig.state_compar} shows the \acp{RMSE} of the estimated sensor trajectories for four SLAM algorithms across four scenarios, compared to theoretical bounds. We observe that the proposed algorithm’s bounds are approximately 30\% lower than those of the filter-based algorithms. This difference arises because the proposed algorithm focus on the posterior $f(\boldsymbol{s}_{0:K},\mathcal{X}|\mathcal{Z}_{1:K} )$, which incorporates all measurements. In contrast, filter-based algorithms work on $f(\boldsymbol{s}_{k},\mathcal{X}|\mathcal{Z}_{1:k} )$ for $k\in\{1,\cdots,K\}$, conditioned only on measurements up to the current time step, resulting in higher bounds. For all algorithms, the bounds are higher in high-process noise scenarios (Scenarios III and IV) compared to low-process noise scenarios (Scenarios I and II). This is due to the PCRB considering the transition density; lower process noise, indicating a more accurate motion model, brings more posterior information and results in lower bounds. Therefore, all algorithms perform better in low process noise scenarios. Furthermore, all algorithms exhibit slightly worse performance in high clutter scenarios (Scenarios II and IV) compared to low clutter scenarios (Scenarios I and III), as the bars are higher. This decline is attributed to the presence of closely spaced clutter measurements in high clutter scenarios, which leads to false alarms and negatively impacts overall performance.

Among the four algorithms, the proposed algorithm demonstrates superior performance due to its batch processing approach, as evidenced by the blue bars being the lowest in Fig.~\ref{Fig.state_compar}. Additionally, the proposed algorithm is robust to both high clutter and high process noise, maintaining close-bound performance in all scenarios, which is indicated by the blue bars being very close to solid black lines in Fig.~\ref{Fig.state_compar}. The robustness is due to the effective solution to the DA for the entire measurement batch and the joint optimization conditioned on the resulting DAs, allowing the algorithm to track all cross-correlations between the sensor trajectory and the map. Among filter-based algorithms, the EK-PMBM SLAM filter, which drops cross-correlations in computation, suffers from information loss, while the RBP-PHD and RBP-PHD2 SLAM filters retain cross-correlations through particles, requiring a sufficient number of particles for good performance. Consequently, the EK-PMBM SLAM filter performs the worst among the algorithms, when a sufficient number of particles are used for the two RBP-based algorithms, as the red bars are the highest in low process noise scenarios. However, 1000 particles are insufficient for the RBP-PHD SLAM filter in high process noise scenarios, leading to worse positioning performance for the RBP-PHD SLAM filter compared to the EK-PMBM SLAM filter, as reflected by the yellow bars being generally highest in high process noise scenarios. The RBP-PHD2 SLAM filter has close-bound performance as 1000 particles are sufficient, but it still underperforms to the proposed algorithm, due to its inherently higher bounds as a filter-based algorithm.

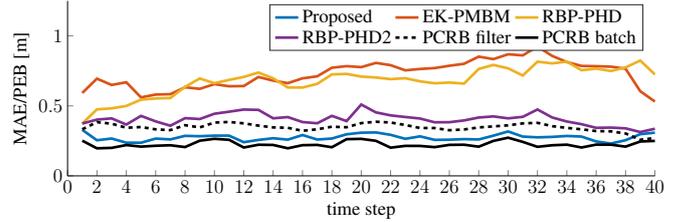
\begin{figure}
\center
\input{PCRB}
\caption{Comparison of RMSE on sensor position estimates changing with time among four algorithms and two bounds for Scenario IV.}
\label{Fig.RCRB_compare}
 \end{figure}

Fig.~\ref{Fig.RCRB_compare} demonstrates that the proposed algorithm consistently outperforms filter-based algorithms in Scenario IV, as the blue line consistently lies below the red, yellow, and purple lines, highlighting the efficacy of the proposed algorithm. Moreover, the proposed algorithm's bound remains stable throughout the trajectory, in contrast to the decreasing trend observed for filter-based bounds, as the solid black line remains stable, while the solid dashed line decreases in general. This stability arises from the batch-processing bounds conditioning on all measurements, unlike filter-based bound, which are conditioned only on measurements up to the current time step.  As time progresses, more measurements can be incorporated, leading to improved performance.

\subsubsection{Mapping Performance}
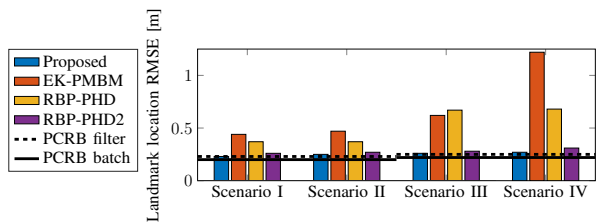
\begin{figure}
\center
\input{compar_bar_mapping}
\caption{Comparison of landmark estimations for 4 algorithms under 4 scenarios.}
\label{Fig.state_compar_mapping}
 \end{figure}

Fig.~\ref{Fig.state_compar_mapping} shows the RMSE of estimated landmark locations for four SLAM algorithms across different scenarios, compared to their respective bounds. We observe that the bounds for batch processing are lower than the bounds for filter-based algorithms, indicated by the solid black lines being lower than the dashed black lines in Fig.~\ref{Fig.state_compar_mapping}. This is because the batch processing incorporates the entire sensor trajectory into the \ac{PIM}, compared to filter-based algorithms that only have snapshots of the sensor state in the bound computation. Consequently, batch processing yields lower bounds than filter-based methods even when all measurements are conditioned. In low process noise scenarios, all bounds are lower due to the more accurate transition model, which also benefits landmark state estimation.

The proposed method is robust to both high clutter and high process noise, and performs the best in landmark state estimation, as evidenced by the blue bars being close to solid black lines and lowest among four algorithms in all four scenarios. The superior performance and the robustness are attributed to the batch processing of the proposed method. In contrast, the EK-PMBM and RBP-PHD SLAM filters perform poorly. The EK-PMBM SLAM filter suffers from information loss, and the RBP-PHD SLAM does not utilize sufficient particles. Their performances degrade further in Scenario IV, due to the challenges posed by high clutter and high process noise for filter-based algorithms. The RBP-PHD2 filter, with sufficient particles, is also robust to high clutter and process noise but still underperforms compared to the proposed algorithm, due to its inherently filter-based processing.

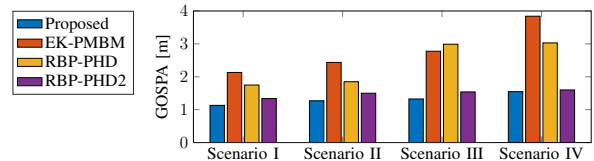
\begin{figure}
\center
\input{compare_GOSPA_mapping}
\caption{Comparison of GOSPA distance for 4 algorithms under 4 scenarios.}
\label{Fig.GOSPA}
 \end{figure}

Fig.~\ref{Fig.GOSPA} shows the GOSPA distance for the four algorithms across four scenarios. Consistent with previous results, the proposed algorithm exhibits the best performance, providing better landmark estimations with fewer false alarms and misdetections, as the blue bars are the lowest. This superior performance is due to the effectiveness of the proposed DA solution for measurement batch and joint optimization in the SLAM results. Among the filter-based algorithms, the RBP-PHD2 filter performs the best due to its use of sufficient samples, where DA problem is solved effectively, resulting in fewer false alarms and misdetections. In contrast, the RBP-PHD and EK-PMBM SLAM filters perform poorly due to insufficient particles or information loss from marginalization, leading to more false alarms and misdetections compared to the RBP-PHD2 SLAM filter.

\section{Conclusions} \label{Sec:conclusion}
This paper presents a novel Graph PMBM-SLAM algorithm, which firstly bridges the \ac{RFS} theory and graph-based SLAM together. By modeling the measurements and the landmarks as \acp{RFS}, a sampling-based algorithm, which combines the Gibbs sampling algorithm and the \ac{MH} algorithm together, is proposed to solve the \ac{DA} problem of all measurements given a sensor trajectory. The GraphSLAM algorithm is applied to estimate the best fit of the sensor trajectory and the map to the joint posterior of the sensor trajectory and the map conditioned on the resulting \ac{DA}. The proposed framework iterates within these two steps until reaching a maximal number of iterations. The marginalization step to merge the SLAM resulting from iterations serves as the post-processing step to approximate the correct joint posterior, where the map is modeled as a \ac{RFS} instead of a list of random vectors. Analysis was carried out in four simulated scenarios through MC simulations. Results demonstrated that the proposed framework can address the \ac{DA} problem of all measurements accurately. Our results also demonstrated the close-to-bound performance of the proposed framework in mapping and positioning, as well as its accuracy and robustness in high clutter and high process noise scenarios. Future work will include the extension to extended object models, and the evaluation with other experimental data sets such as visual and lidar data sets.

\balance 
\bibliography{IEEEabrv,Bibliography}

\newpage
\pagestyle{empty}
\beginsupplement
\makeatletter

\noindent \large{\textbf{Supplemental material: ``Batch SLAM with PMBM Data Association Sampling and Graph-Based Optimization''}}
\normalsize
\allowdisplaybreaks

\input{Supplementary}

\end{document}

%% file: acronyms.tex
\acrodef{MMSE}{Minimum Mean Squared Error}

\acrodef{MSE}{mean square error}

\acrodef{PSD}{power spectral density}

\acrodef{RMSE}{root mean squared error}
\acrodef{SLR}{statistical linear regression}

\acrodef{IPLF}{iterated posterior linearization filter}

\acrodef{ue}[UE]{user equipment}
\acrodef{bs}[BS]{base station}
\acrodef{va}[VA]{virtual anchor}
\acrodef{sp}[SP]{scattering  point}
\acrodef{fov}[FoV]{field-of-view}   
\acrodef{los}[LOS]{line-of-sight}
\acrodef{nlos}[NLOS]{non-line-of-sight}
\acrodef{PMBM}[PMBM]{Poisson  multi-Bernoulli  mixture}
\acrodef{PMB(M)}[PMB(M)]{Poisson  multi-Bernoulli  (mixture)}
\acrodef{PMB}[PMB]{Poisson  multi-Bernoulli}
\acrodef{RFS}[RFS]{random finite set}
\acrodef{PPP}[PPP]{Poisson point process}
\acrodef{MBM}[MBM]{multi-Bernoulli  mixture}
\acrodef{MB}[MB]{multi-Bernoulli}

\acrodef{ekf}[EKF]{extended Kalman filter}
\acrodef{PDF}[PDF]{probability density function}

\acrodef{ckf}[CKF]{cubature Kalman filter}
\acrodef{rbp}[RBP]{Rao-Blackwellized particle}
\acrodef{gospa}[GOSPA]{generalized optimal subpattern assignment}
\acrodef{SLAM}[SLAM]{simultaneous localization and mapping}

\acrodef{TOA}[TOA]{time of arrival}
\acrodef{AOA}[AOA]{angles of arrival}
\acrodef{AOD}[AOD]{angles of departure}

\acrodef{IF}[IF]{information filter}
\acrodef{EIF}[EIF]{extended information filter}
\acrodef{kf}[KF]{Kalman filter}
\acrodef{PMF}[PMF]{probability mass function}
\acrodef{MAP}[MAP]{maximum a posteriori estimation}

\acrodef{DA}[DA]{data association}
\acrodef{OFDM}[OFDM]{Orthogonal Frequency Division Multiplexing}

\acrodef{PCRB}[PCRB]{posterior Cram{\'e}r-Rao bound}

\acrodef{EM}[EM]{expectation-maximization}
\acrodef{FOV}[FOV]{field of view}
\acrodef{RB}[RB]{Rao-Blackwellized }

\acrodef{ISAC}[ISAC]{integrated sensing and communication}

\acrodef{EK}[EK]{extended Kalman}
\acrodef{EKF}[EKF]{extended Kalman filter}

\acrodef{LMB}[LMB]{labeled multi-Bernoulli}
\acrodef{GLMB}[$\delta$-GLMB]{$\delta$-generalized labeled multi-Bernoulli}
\acrodef{PHD}[PHD]{probability hypothesis density}

\acrodef{OID}[OID]{optimal importance density}

\acrodef{MTT}[MTT]{multi-target tracking}

\acrodef{MCMC}[MCMC]{Markov chain Monte Carlo}

\acrodef{MH}[MH]{Metropolis-Hastings}

\acrodef{NMI}[NMI]{normalized mutual information}

\acrodef{PCRB}[PCRB]{posterior Cram{\'e}r-Rao bound}

\acrodef{PIM}[PIM]{posterior information matrix}

\acrodef{OID}[OID]{optimal importance density}

\acrodef{MAE}[MAE]{mean absoluate error} 

\acrodef{KL}[KL]{Kullback–Leibler}

%% file: compar_bar_pos.tex
%
%
\definecolor{mycolor1}{rgb}{0.00000,0.44700,0.74100}%
\definecolor{mycolor2}{rgb}{0.85000,0.32500,0.09800}%
\definecolor{mycolor3}{rgb}{0.92900,0.69400,0.12500}%
\definecolor{mycolor4}{rgb}{0.49400,0.18400,0.55600}
\definecolor{mycolor5}{rgb}{0,0,0}%
\begin{tikzpicture}[scale=0.85\linewidth/14cm]

\begin{axis}[%
width=3.842in,
height=1.281in,
at={(3.465in,2.378in)},
scale only axis,
bar shift auto,
xmin=0.5,
xmax=4.5,
xtick={1,2,3,4},
xticklabels={{Scenario I},{Scenario II},{Scenario III},{Scenario IV}},
ticklabel style={font=\large},
ymin=0,
ymax=0.8,
ylabel style={font=\color{white!15!black},font=\large},
ylabel={UE position RMSE [m]},
axis background/.style={fill=white},
legend style={at={(-0.15,1)}, anchor=north east, legend cell align=left, align=left, draw=white!15!black,font=\large}
]
\addplot[ybar, bar width=0.145, fill=mycolor1, draw=black, area legend] table[row sep=crcr] {%
1	0.21\\
2	0.23\\
3	0.28\\
4	0.29\\ 
};
\addplot[forget plot, color=white!15!black] table[row sep=crcr] {%
0.5	0\\
3.5	0\\
};
\addlegendentry{Proposed}

\addplot[ybar, bar width=0.145, fill=mycolor2, draw=black, area legend] table[row sep=crcr] {%
1	0.45\\
2	0.46\\
3	0.61\\
4	0.69\\
};
\addplot[forget plot, color=white!15!black] table[row sep=crcr] {%
0.5	0\\
3.5	0\\
};
\addlegendentry{EK-PMBM}

\addplot[ybar, bar width=0.145, fill=mycolor3, draw=black, area legend] table[row sep=crcr] {%
1	0.39\\
2	0.38\\
3	0.69\\
4	0.69\\
};
\addplot[forget plot, color=white!15!black] table[row sep=crcr] {%
0.5	0\\
3.5	0\\
};
\addlegendentry{RBP-PHD}

\addplot[ybar, bar width=0.145, fill=mycolor4, draw=black, area legend] table[row sep=crcr] {%
1	0.32\\
2	0.32\\
3	0.38\\
4	0.41\\
};
\addplot[forget plot, color=white!15!black] table[row sep=crcr] {%
0.5	0\\
3.5	0\\
};
\addlegendentry{RBP-PHD2}

\addplot [color=mycolor5, dashed,line width=2.0pt]
  table[row sep=crcr]{%
0.5	0.29\\
2.5	0.29\\
};
\addlegendentry{PCRB filter}

\addplot [color=mycolor5, line width=2.0pt]
  table[row sep=crcr]{%
0.5	0.19\\
2.5	0.19\\
};
\addlegendentry{PCRB batch}

\addplot [color=mycolor5, dashed,line width=2.0pt, forget plot]
  table[row sep=crcr]{%
2.5	0.35\\
4.5	0.35\\
};

\addplot [color=mycolor5, line width=2.0pt, forget plot]
  table[row sep=crcr]{%
2.5	0.24\\
4.5	0.24\\
};

\end{axis}
\end{tikzpicture}%

%% file: compar_bar_heading.tex
%
%
\definecolor{mycolor1}{rgb}{0.00000,0.44700,0.74100}%
\definecolor{mycolor2}{rgb}{0.85000,0.32500,0.09800}%
\definecolor{mycolor3}{rgb}{0.92900,0.69400,0.12500}%
\definecolor{mycolor4}{rgb}{0.49400,0.18400,0.55600}
\definecolor{mycolor5}{rgb}{0,0,0}%
\begin{tikzpicture}[scale=0.85\linewidth/14cm]

\begin{axis}[%
width=3.842in,
height=1.281in,
at={(3.465in,2.378in)},
scale only axis,
bar shift auto,
xmin=0.5,
xmax=4.5,
xtick={1,2,3,4},
xticklabels={{Scenario I},{Scenario II},{Scenario III},{Scenario IV}},
ticklabel style={font=\large},
ymin=0,
ymax=0.5,
ylabel style={font=\color{white!15!black},font=\large},
ylabel={UE heading RMSE [deg]},
axis background/.style={fill=white},
legend style={at={(-0.15,1)}, anchor=north east, legend cell align=left, align=left, draw=white!15!black,font=\large}
]
\addplot[ybar, bar width=0.145, fill=mycolor1, draw=black, area legend] table[row sep=crcr] {%
1	0.11\\
2	0.11\\
3	0.19\\
4	0.19\\ 
};
\addplot[forget plot, color=white!15!black] table[row sep=crcr] {%
0.5	0\\
3.5	0\\
};
\addlegendentry{Proposed}

\addplot[ybar, bar width=0.145, fill=mycolor2, draw=black, area legend] table[row sep=crcr] {%
1	0.25\\
2	0.27\\
3	0.35\\
4	0.40\\
};
\addplot[forget plot, color=white!15!black] table[row sep=crcr] {%
0.5	0\\
3.5	0\\
};
\addlegendentry{EK-PMBM}

\addplot[ybar, bar width=0.145, fill=mycolor3, draw=black, area legend] table[row sep=crcr] {%
1	0.21\\
2	0.23\\
3	0.47\\
4	0.49\\
};
\addplot[forget plot, color=white!15!black] table[row sep=crcr] {%
0.5	0\\
3.5	0\\
};
\addlegendentry{RBP-PHD}

\addplot[ybar, bar width=0.145, fill=mycolor4, draw=black, area legend] table[row sep=crcr] {%
1	0.17\\
2	0.19\\
3	0.26\\
4	0.29\\
};
\addplot[forget plot, color=white!15!black] table[row sep=crcr] {%
0.5	0\\
3.5	0\\
};
\addlegendentry{RBP-PHD2}

\addplot [color=mycolor5, dashed,line width=2.0pt]
  table[row sep=crcr]{%
0.5	0.17\\
2.5	0.17\\
};
\addlegendentry{PCRB filter}

\addplot [color=mycolor5, line width=2.0pt]
  table[row sep=crcr]{%
0.5	0.11\\
2.5	0.11\\
};
\addlegendentry{PCRB batch}

\addplot [color=mycolor5, dashed,line width=2.0pt, forget plot]
  table[row sep=crcr]{%
2.5	0.25\\
4.5	0.25\\
};

\addplot [color=mycolor5, line width=2.0pt, forget plot]
  table[row sep=crcr]{%
2.5	0.18\\
4.5	0.18\\
};

\end{axis}
\end{tikzpicture}%

%% file: compar_bar_bias.tex
%
%
\definecolor{mycolor1}{rgb}{0.00000,0.44700,0.74100}%
\definecolor{mycolor2}{rgb}{0.85000,0.32500,0.09800}%
\definecolor{mycolor3}{rgb}{0.92900,0.69400,0.12500}%
\definecolor{mycolor4}{rgb}{0.49400,0.18400,0.55600}
\definecolor{mycolor5}{rgb}{0,0,0}%
\begin{tikzpicture}[scale=0.85\linewidth/14cm]

\begin{axis}[%
width=3.842in,
height=1.281in,
at={(3.465in,2.378in)},
scale only axis,
bar shift auto,
xmin=0.5,
xmax=4.5,
xtick={1,2,3,4},
xticklabels={{Scenario I},{Scenario II},{Scenario III},{Scenario IV}},
ticklabel style={font=\large},
ymin=0,
ymax=0.5,
ylabel style={font=\color{white!15!black},font=\large},
ylabel={UE clock bias RMSE [m]},
axis background/.style={fill=white},
legend style={at={(-0.15,1)}, anchor=north east, legend cell align=left, align=left, draw=white!15!black,font=\large}
]
\addplot[ybar, bar width=0.145, fill=mycolor1, draw=black, area legend] table[row sep=crcr] {%
1	0.13\\
2	0.13\\
3	0.17\\
4	0.18\\ 
};
\addplot[forget plot, color=white!15!black] table[row sep=crcr] {%
0.5	0\\
3.5	0\\
};
\addlegendentry{Proposed}

\addplot[ybar, bar width=0.145, fill=mycolor2, draw=black, area legend] table[row sep=crcr] {%
1	0.24\\
2	0.25\\
3	0.40\\
4	0.45\\
};
\addplot[forget plot, color=white!15!black] table[row sep=crcr] {%
0.5	0\\
3.5	0\\
};
\addlegendentry{EK-PMBM}

\addplot[ybar, bar width=0.145, fill=mycolor3, draw=black, area legend] table[row sep=crcr] {%
1	0.25\\
2	0.24\\
3	0.44\\
4	0.44\\
};
\addplot[forget plot, color=white!15!black] table[row sep=crcr] {%
0.5	0\\
3.5	0\\
};
\addlegendentry{RBP-PHD}

\addplot[ybar, bar width=0.145, fill=mycolor4, draw=black, area legend] table[row sep=crcr] {%
1	0.20\\
2	0.20\\
3	0.25\\
4	0.27\\
};
\addplot[forget plot, color=white!15!black] table[row sep=crcr] {%
0.5	0\\
3.5	0\\
};
\addlegendentry{RBP-PHD2}

\addplot [color=mycolor5, dashed,line width=2.0pt]
  table[row sep=crcr]{%
0.5	0.17\\
2.5	0.17\\
};
\addlegendentry{PCRB filter}

\addplot [color=mycolor5, line width=2.0pt]
  table[row sep=crcr]{%
0.5	0.10\\
2.5	0.10\\
};
\addlegendentry{PCRB batch}

\addplot [color=mycolor5, dashed,line width=2.0pt, forget plot]
  table[row sep=crcr]{%
2.5	0.21\\
4.5	0.21\\
};

\addplot [color=mycolor5, line width=2.0pt, forget plot]
  table[row sep=crcr]{%
2.5	0.14\\
4.5	0.14\\
};

\end{axis}
\end{tikzpicture}%

%% file: PCRB.tex
%
%
\definecolor{mycolor1}{rgb}{0.00000,0.44700,0.74100}%
\definecolor{mycolor2}{rgb}{0.85000,0.32500,0.09800}%
\definecolor{mycolor3}{rgb}{0.92900,0.69400,0.12500}%
\definecolor{mycolor4}{rgb}{0.49400,0.18400,0.55600}
\definecolor{mycolor5}{rgb}{0,0,0}%
\begin{tikzpicture}[scale=0.8\linewidth/14cm]

\begin{axis}[%
width=6.028in,
height=1.809in,
at={(1.011in,2.014in)},
scale only axis,
xmin=0,
xmax=40,
xlabel style={font=\color{white!15!black},font=\Large},
xlabel={time step},
ticklabel style={font=\Large},
ymin=0,
ymax=1.25,
yminorticks=true,
ylabel style={font=\color{white!15!black},font=\Large},
ylabel={MAE/PEB [m]},
axis background/.style={fill=white},
axis x line*=bottom,
axis y line*=left,
legend style={legend cell align=left, align=left, draw=white!15!black,font=\Large,legend columns=3}
]

\addplot [color=mycolor1, line width=2.0pt]
table[row sep=crcr]{%
1	0.329115798750277\\
2	0.25450765052488\\
3	0.266792530121479\\
4	0.237102957263112\\
5	0.236862878468914\\
6	0.266475108042213\\
7	0.260502058971092\\
8	0.286065011927385\\
9	0.283394095472154\\
10	0.287623417779019\\
11	0.287669360265205\\
12	0.240703118145424\\
13	0.256813063617578\\
14	0.268897013103315\\
15	0.258913710561553\\
16	0.291923995934201\\
17	0.259654717534275\\
18	0.266869148349839\\
19	0.297078896611332\\
20	0.308768477207626\\
21	0.310351774260241\\
22	0.293321227229279\\
23	0.264707814379237\\
24	0.282399466902986\\
25	0.257396220795871\\
26	0.258503209345715\\
27	0.26258038872981\\
28	0.260227107999406\\
29	0.287475322826636\\
30	0.3173877794565\\
31	0.28078777322624\\
32	0.275384533526158\\
33	0.278442962042752\\
34	0.28514434360295\\
35	0.280831210352693\\
36	0.246065723425138\\
37	0.230898581791878\\
38	0.254850608882532\\
39	0.297905982994521\\
40	0.308786362783943\\
};
\addlegendentry{Proposed}

\addplot [color=mycolor2, line width=2.0pt]
  table[row sep=crcr]{%
  1	0.592329091824408\\
2	0.694681738700614\\
3	0.649703386477434\\
4	0.669016009579981\\
5	0.560735904350104\\
6	0.581226353309038\\
7	0.583783539771509\\
8	0.633311858394694\\
9	0.621904784581231\\
10	0.655626421741157\\
11	0.640446196759431\\
12	0.64190672573224\\
13	0.706684750204142\\
14	0.68071292337803\\
15	0.662512887898333\\
16	0.697051676422405\\
17	0.711357021584589\\
18	0.772950374928163\\
19	0.784096451805851\\
20	0.777127334298392\\
21	0.807331780722879\\
22	0.791055558683666\\
23	0.754271693312575\\
24	0.763794783510083\\
25	0.770428838721113\\
26	0.786905904112921\\
27	0.800977871050762\\
28	0.852052717703908\\
29	0.834362607663019\\
30	0.868946378065818\\
31	0.861013937287324\\
32	0.924291386524485\\
33	0.857495766054371\\
34	0.813208582817518\\
35	0.785578688078645\\
36	0.783733815561572\\
37	0.780431533584956\\
38	0.766045312740655\\
39	0.605044748564125\\
40	0.531048577538176\\
};
\addlegendentry{EK-PMBM}

\addplot [color=mycolor3, line width=2.0pt]
  table[row sep=crcr]{%
1 0.373863145114874 \\ 
2 0.475201136771880 \\ 
3 0.482399211676562 \\ 
4 0.499837825969880 \\ 
5 0.544687119634598 \\ 
6 0.553198717166770 \\ 
7 0.556302216327402 \\ 
8 0.636193658749928 \\ 
9 0.695933319260833 \\ 
10 0.662351849371729 \\ 
11 0.685585969813398 \\ 
12 0.706982833707582 \\ 
13 0.738465920509285 \\ 
14 0.697188978128301 \\ 
15 0.631656744931131 \\ 
16 0.631355998986593 \\ 
17 0.657458457606198 \\ 
18 0.725126439908584 \\ 
19 0.728620311255564 \\ 
20 0.709512326289621 \\ 
21 0.702833640765092 \\ 
22 0.691460928055581 \\ 
23 0.697766275148566 \\ 
24 0.675659815042117 \\ 
25 0.661613540480194 \\ 
26 0.666887886357519 \\ 
27 0.659440289481309 \\ 
28 0.764179975474465 \\ 
29 0.792755947757169 \\ 
30 0.768554886851691 \\ 
31 0.717013253015237 \\ 
32 0.816135077901149 \\ 
33 0.803854095499970 \\ 
34 0.816430060978351 \\ 
35 0.754981396400060 \\ 
36 0.766907532117125 \\ 
37 0.749336809637518 \\ 
38 0.774830741851127 \\ 
39 0.824122112137825 \\ 
40 0.724522383760577 \\ 
};
\addlegendentry{RBP-PHD}

\addplot [color=mycolor4, line width=2.0pt]
  table[row sep=crcr]{%
1 0.373863145114874 \\ 
2 0.402027244576322 \\ 
3 0.410742732646126 \\ 
4 0.366864678248508 \\ 
5 0.428681300225854 \\ 
6 0.389360225404136 \\ 
7 0.358412773081595 \\ 
8 0.412442458126769 \\ 
9 0.406072130755841 \\ 
10 0.443899625765266 \\ 
11 0.458594706758836 \\ 
12 0.474022159747056 \\ 
13 0.471569847871372 \\ 
14 0.411353838876599 \\ 
15 0.420084929297364 \\ 
16 0.384521210137945 \\ 
17 0.375946826821379 \\ 
18 0.429045842995462 \\ 
19 0.391263085795651 \\ 
20 0.509749314128948 \\ 
21 0.453556283883019 \\ 
22 0.432384431487202 \\ 
23 0.420688621310665 \\ 
24 0.409979091980253 \\ 
25 0.382905292431813 \\ 
26 0.383512772799884 \\ 
27 0.396093625446079 \\ 
28 0.416914629571410 \\ 
29 0.425010260931833 \\ 
30 0.411076257121486 \\ 
31 0.421917901925686 \\ 
32 0.474440060414111 \\ 
33 0.417389884511459 \\ 
34 0.388569404706714 \\ 
35 0.369459052272692 \\ 
36 0.343324634921594 \\ 
37 0.345065643996477 \\ 
38 0.339719545554270 \\ 
39 0.313410782703606 \\ 
40 0.336737421246210 \\ 
};
\addlegendentry{RBP-PHD2}

\addplot [color=mycolor5, dashed, line width=2.0pt]
  table[row sep=crcr]{%
1 0.332877834411985 \\ 
2 0.383673601829218 \\ 
3 0.373028304724987 \\ 
4 0.343760634137552 \\ 
5 0.349014004158465 \\ 
6 0.332300260206219 \\ 
7 0.326867039719854 \\ 
8 0.362212201393296 \\ 
9 0.346289615998398 \\ 
10 0.380883913534619 \\ 
11 0.385189143852696 \\ 
12 0.375386942864258 \\ 
13 0.359100304143953 \\ 
14 0.346374693715548 \\ 
15 0.343511792231103 \\ 
16 0.321997210506032 \\ 
17 0.332809052729327 \\ 
18 0.351251816472301 \\ 
19 0.347380372316787 \\ 
20 0.376965051634130 \\ 
21 0.386899704429813 \\ 
22 0.381285254041230 \\ 
23 0.363513924897195 \\ 
24 0.342105709149904 \\ 
25 0.341829934779741 \\ 
26 0.326220768997376 \\ 
27 0.331480012955888 \\ 
28 0.346140697896223 \\ 
29 0.353566437521872 \\ 
30 0.361071395198212 \\ 
31 0.374965365691035 \\ 
32 0.379022309637607 \\ 
33 0.355299381675252 \\ 
34 0.342905796990360 \\ 
35 0.332668469613309 \\ 
36 0.320170787258281 \\ 
37 0.318140878944951 \\ 
38 0.302921587218733 \\ 
39 0.257920255619278 \\ 
40 0.270722040629413 \\
};
\addlegendentry{PCRB filter}

\addplot [color=mycolor5, line width=2.0pt]
  table[row sep=crcr]{%
1	0.251578101968872\\
2	0.196911160887535\\
3	0.200039998715574\\
4	0.220108827287418\\
5	0.208987218366362\\
6	0.21519220604827\\
7	0.218730931336664\\
8	0.205775072678894\\
9	0.251345036979643\\
10	0.264842805605618\\
11	0.257829840568579\\
12	0.202572278279789\\
13	0.22250417595858\\
14	0.221011359271969\\
15	0.197096629237742\\
16	0.21935160095506\\
17	0.221268587223097\\
18	0.204560939202037\\
19	0.262312097866767\\
20	0.264356914910146\\
21	0.250397898530938\\
22	0.201461876709764\\
23	0.214338189366531\\
24	0.214010033683072\\
25	0.2053385028435\\
26	0.221578618703886\\
27	0.222605414630448\\
28	0.208585241670946\\
29	0.249925808158278\\
30	0.27284737850866\\
31	0.242833497353108\\
32	0.207556609875471\\
33	0.218731502860456\\
34	0.223043834860949\\
35	0.202698339233795\\
36	0.223411965139372\\
37	0.222984048330968\\
38	0.208906522945507\\
39	0.245472910869289\\
40	0.249861837746596\\
};
\addlegendentry{PCRB batch}

\end{axis}
\end{tikzpicture}%

%% file: compar_bar_mapping.tex
%
%
\definecolor{mycolor1}{rgb}{0.00000,0.44700,0.74100}%
\definecolor{mycolor2}{rgb}{0.85000,0.32500,0.09800}%
\definecolor{mycolor3}{rgb}{0.92900,0.69400,0.12500}%
\definecolor{mycolor4}{rgb}{0.49400,0.18400,0.55600}
\definecolor{mycolor5}{rgb}{0,0,0}%
\begin{tikzpicture}[scale=0.85\linewidth/14cm]

\begin{axis}[%
width=3.842in,
height=1.281in,
at={(3.465in,2.378in)},
scale only axis,
bar shift auto,
xmin=0.5,
xmax=4.5,
xtick={1,2,3,4},
xticklabels={{Scenario I},{Scenario II},{Scenario III},{Scenario IV}},
ticklabel style={font=\large},
ymin=0,
ymax=1.25,
ylabel style={font=\color{white!15!black},font=\large},
ylabel={Landmark location RMSE [m]},
axis background/.style={fill=white},
legend style={at={(-0.15,1)}, anchor=north east, legend cell align=left, align=left, draw=white!15!black,font=\large}
]
\addplot[ybar, bar width=0.145, fill=mycolor1, draw=black, area legend] table[row sep=crcr] {%
1	0.23\\
2	0.25\\
3	0.26\\
4	0.27\\ 
};
\addplot[forget plot, color=white!15!black] table[row sep=crcr] {%
0.5	0\\
3.5	0\\
};
\addlegendentry{Proposed}

\addplot[ybar, bar width=0.145, fill=mycolor2, draw=black, area legend] table[row sep=crcr] {%
1	0.44\\
2	0.47\\
3	0.62\\
4	1.22\\
};
\addplot[forget plot, color=white!15!black] table[row sep=crcr] {%
0.5	0\\
3.5	0\\
};
\addlegendentry{EK-PMBM}

\addplot[ybar, bar width=0.145, fill=mycolor3, draw=black, area legend] table[row sep=crcr] {%
1	0.37\\
2	0.37\\
3	0.67\\
4	0.68\\
};
\addplot[forget plot, color=white!15!black] table[row sep=crcr] {%
0.5	0\\
3.5	0\\
};
\addlegendentry{RBP-PHD}

\addplot[ybar, bar width=0.145, fill=mycolor4, draw=black, area legend] table[row sep=crcr] {%
1	0.26\\
2	0.27\\
3	0.28\\
4	0.31\\
};
\addplot[forget plot, color=white!15!black] table[row sep=crcr] {%
0.5	0\\
3.5	0\\
};
\addlegendentry{RBP-PHD2}

\addplot [color=mycolor5, dashed,line width=2.0pt]
  table[row sep=crcr]{%
0.5	0.23\\
2.5	0.23\\
};
\addlegendentry{PCRB filter}

\addplot [color=mycolor5, line width=2.0pt]
  table[row sep=crcr]{%
0.5	0.20\\
2.5	0.20\\
};
\addlegendentry{PCRB batch}

\addplot [color=mycolor5, dashed,line width=2.0pt, forget plot]
  table[row sep=crcr]{%
2.5	0.25\\
4.5	0.25\\
};

\addplot [color=mycolor5, line width=2.0pt, forget plot]
  table[row sep=crcr]{%
2.5	0.22\\
4.5	0.22\\
};

\end{axis}
\end{tikzpicture}%

%% file: compare_GOSPA_mapping.tex
%
%
\definecolor{mycolor1}{rgb}{0.00000,0.44700,0.74100}%
\definecolor{mycolor2}{rgb}{0.85000,0.32500,0.09800}%
\definecolor{mycolor3}{rgb}{0.92900,0.69400,0.12500}%
\definecolor{mycolor4}{rgb}{0.49400,0.18400,0.55600}
\definecolor{mycolor5}{rgb}{0,0,0}%
\begin{tikzpicture}[scale=0.85\linewidth/14cm]

\begin{axis}[%
width=3.842in,
height=1.281in,
at={(3.465in,2.378in)},
scale only axis,
bar shift auto,
xmin=0.5,
xmax=4.5,
xtick={1,2,3,4},
xticklabels={{Scenario I},{Scenario II},{Scenario III},{Scenario IV}},
ymin=0,
ymax=4,
ticklabel style={font=\large},
ylabel style={font=\color{white!15!black},font=\large},
ylabel={GOSPA [m]},
axis background/.style={fill=white},
legend style={at={(-0.15,1)}, anchor=north east, legend cell align=left, align=left, draw=white!15!black,font=\large}
]
\addplot[ybar, bar width=0.145, fill=mycolor1, draw=black, area legend] table[row sep=crcr] {%
1	1.1320\\
2	1.2709\\
3	1.3259\\
4	1.5483\\
};
\addplot[forget plot, color=white!15!black] table[row sep=crcr] {%
0.5	0\\
3.5	0\\
};
\addlegendentry{Proposed}

\addplot[ybar, bar width=0.145, fill=mycolor2, draw=black, area legend] table[row sep=crcr] {%
1	2.1318\\
2	2.4365\\
3	2.7762\\
4	3.841\\
};
\addplot[forget plot, color=white!15!black] table[row sep=crcr] {%
0.5	0\\
3.5	0\\
};
\addlegendentry{EK-PMBM}

\addplot[ybar, bar width=0.145, fill=mycolor3, draw=black, area legend] table[row sep=crcr] {%
1	1.75\\
2	1.85\\
3	2.99\\
4	3.03\\
};
\addplot[forget plot, color=white!15!black] table[row sep=crcr] {%
0.5	0\\
3.5	0\\
};
\addlegendentry{RBP-PHD}

\addplot[ybar, bar width=0.145, fill=mycolor4, draw=black, area legend] table[row sep=crcr] {%
1	1.34\\
2	1.50\\
3	1.54\\
4	1.60\\
};
\addplot[forget plot, color=white!15!black] table[row sep=crcr] {%
0.5	0\\
3.5	0\\
};
\addlegendentry{RBP-PHD2}

\end{axis}
\end{tikzpicture}%

%% file: Supplementary.tex
\appendices
\section{Proof of Theorem 1} \label{prove_PMBM}

\newcommand{\Ucal}{\mathcal{U}}
\newcommand{\Xcal}{\mathcal{X}}
\newcommand{\Ycal}{\mathcal{Y}}
\newcommand{\Zcal}{\mathcal{Z}}
\newcommand{\sbf}{\boldsymbol{s}}
\newcommand{\xbf}{\boldsymbol{x}}
\newcommand{\ybf}{\boldsymbol{y}}
\newcommand{\zbf}{\boldsymbol{z}}

The proof of Theorem~\ref{theo:posterior} relies on the following lemma. 

\begin{lemma}
    \label{lemma:BatchLikelihood}

The likelihood of the sequence of measurements can be expressed as
\begin{align}
&g(\mathcal{Z}_{1:K} \mid \boldsymbol{s}_{1:K},\mathcal{X} ) = e^{-K\int c(\boldsymbol{z}) \text{d} \boldsymbol{z}} \label{eq:jointlikelihood}
\\ & \sum_{j\in \mathbb{J}}
\sum_{\mathcal{X}_{\mathrm{U}}\biguplus\mathcal{Y}^{j,1}\biguplus\dots\biguplus\mathcal{Y}^{j,|\mathbb{I}^{j}|}=\mathcal{X}}
\prod_{\boldsymbol{x} \in \mathcal{X}_{\mathrm{U}}}p_{\text{U}}(\boldsymbol{x},\boldsymbol{s}_{1:K}) \nonumber
\\ &  
\prod_{i=1}^{|\mathbb{I}^{j}|}t(\mathcal{Z}_{1:K}^{j,i}|\boldsymbol{s}_{1:K},\mathcal{Y}^{j,i}).\nonumber
\end{align}
Here, $\mathcal{Y}^{j,1}\biguplus\dots\biguplus\mathcal{Y}^{j,|\mathbb{I}^{j}|}$ are all detected landmarks, and $p_{\text{U}}(\boldsymbol{x},\boldsymbol{s}_{1:K})=\prod_{k=1}^{K}(1-p_{\text{D}}(\boldsymbol{x},\boldsymbol{s}_{k}))$ denotes the misdetection probability for landmarks that have not been detected for the whole time-period. Moreover,  $t(\mathcal{Z}_{1:K}^{j,i}|\boldsymbol{s}_{1:K},\mathcal{Y}^{j,i})$ denotes the  likelihood of $\mathcal{Z}_{1:K}^{j,i}$ and is given by \eqref{likelihood_single_tar}.

\end{lemma}

\begin{proof}
    We use induction to prove Lemma~\ref{lemma:BatchLikelihood}. For the base case, $K=1$, there is only one valid partition, $\mathbb{J}=\{1\}$, $|\mathbb{I}^{j}|=|\mathcal{Z} _{1}|$ and $\Zcal_{1:K}^{j,i} =\Zcal_{1:1}^{1,i}= \{\zbf_1^i\}$. We can therefore simplify \eqref{eq:jointlikelihood} to 
    \begin{equation}
    \begin{split}
                &g(\Zcal_1 \mid \sbf_1, \Xcal) = e^{- \int c(\zbf) \mathrm{d} \zbf} 
                \\ & 
\sum_{\mathcal{X}_{\mathrm{U}}\biguplus\mathcal{Y}^{1,1}\biguplus\dots\biguplus\mathcal{Y}^{1,|\mathcal{Z} _{1}|}=\mathcal{X}}
\prod_{\xbf \in \Xcal_{\mathrm{U}}}p_{\text{U}}(\xbf,\boldsymbol{s}_{1:K}) 
\prod_{i=1}^{|\mathcal{Z} _{1}|}t(\{ \zbf_1^i\}|\boldsymbol{s}_1,\mathcal{Y}^{1,i}).
    \end{split}
    \end{equation}
This equation holds since it is equivalent to (13) in \cite{garcia2018poisson}. 

To complete the inductive proof, we show that if \eqref{eq:jointlikelihood} holds for $K$, then we can also express 
\begin{equation}
    g(\mathcal{Z}_{1:K+1} \mid \boldsymbol{s}_{1:K+1},\mathcal{X}) = g(\mathcal{Z}_{1:K} \mid \boldsymbol{s}_{1:K},\mathcal{X})
g(\mathcal{Z}_{K+1} \mid \boldsymbol{s}_{K+1},\mathcal{X})
\end{equation}
on the same form. For completeness, we will now use the notations $\mathbb{J}_K$ and $\mathbb{I}^j_K$ to clarify that these sets depend on $K$. 

To prove the induction we first use \cite[eq.~(25)-(27)]{garcia2018poisson} to write 
\begin{equation}
    \begin{split}
    &g(\mathcal{Z}_{K+1} \mid \boldsymbol{s}_{K+1},\mathcal{X}) = \\
    &
    \sum_{\mathcal{Z}^1_{K+1} \biguplus \dots \biguplus \mathcal{Z}^{ |\mathbb{I}^{j}_K|}_{K+1} \biguplus \mathcal{Z}^y_{K+1} = \Zcal_{K+1} }
    g (\Zcal^y_{K+1} \mid \sbf_{K+1}, \Xcal_\mathrm{U}) \\
    & \prod_{i=1}^{ |\mathbb{I}^{j}_K| } t^{-c} (\Zcal^i_{K+1} \mid \sbf_{K+1}, \Ycal^{j,i}),   \label{eq:K+1-likelihood1}
    \end{split}
\end{equation}
where $t^{-c}$ is identical to $t$ except that we have replaced $c(\zbf)$ with $0$, that is, it assumes that there is no clutter. We note that the above equation sums over all possible assignments of measurements in $\Zcal_{K+1}$ to the previously undetected landmarks  $\Xcal_\mathrm{U}$ (these measurements can also be clutter) and detected landmarks $\Ycal^{j,i}$. Second, we use \cite[eq.~(13)]{garcia2018poisson} to write 
\begin{equation}
    \begin{split}
    &g (\Zcal^y_{K+1} \mid \sbf_{K+1}, \Xcal_\mathrm{U}) = e^{-\int c(\boldsymbol{z}) \text{d} \boldsymbol{z}} \sum_{ \Ucal \biguplus \Ycal_1 \biguplus \dots \biguplus \Ycal_{ |\Zcal^y_{K+1}|} = \Xcal_\mathrm{U}  } \\
    & \prod_{\xbf \in \Ucal} p_{\text{U}}(\xbf,\boldsymbol{s}_{K+1}) \prod_{i=1}^{|\Zcal^y_{K+1}|} \tilde{l}(\zbf^{y,i}_{K+1} \mid \sbf_{K+1}, \Ycal_i), \label{eq:K+1-likelihood2}
    \end{split}
\end{equation}
where 
\begin{align}
    \tilde{l}(\zbf \mid \sbf, \Ycal)=\begin{cases}
        p_\mathrm{D}(\sbf,\xbf)f(\zbf \mid \xbf,\sbf)& \Ycal = \{\xbf\}, \\
        c(\zbf) & \Ycal = \emptyset, \\
        0 & |\Ycal| >1.
    \end{cases}   
\end{align}
Here $\Zcal^y_{K+1}$ denotes the set of measurements that are not generated from previously detected landmarks at time step $K+1$, and $\Xcal_\mathrm{U}$, which is the set of landmarks that are undetected in $\Zcal_{1:K}$, is separated into the landmarks are also undetected at time step $K+1$, denoted $\Ucal$, and (possibly empty) sets of landmarks that gave rise to measurements in $\Zcal^y_{K+1} =\{\zbf^{y,1}_{K+1},\dots,\zbf^{y,|\Zcal^y_{K+1}|}_{K+1}\}$.
Combining \eqref{eq:K+1-likelihood1} and \eqref{eq:K+1-likelihood2} yields
\begin{equation}
    \begin{split}
        &g(\mathcal{Z}_{K+1} \mid \boldsymbol{s}_{K+1},\mathcal{X}) =  e^{-\int c(\boldsymbol{z}) \text{d} \boldsymbol{z}} \sum_{\Zcal^1_{K+1} \biguplus \dots \biguplus Z^{ |\mathbb{I}^{j}_K|}_{K+1} \biguplus \Zcal^y_{K+1} = \Zcal_{K+1} }\\
    &
    \sum_{ \Ucal \biguplus \Ycal_1 \biguplus \dots \biguplus \Ycal_{ |\Zcal^y_{K+1}|} = \Xcal_\mathrm{U}  } \prod_{i=1}^{ |\mathbb{I}^{j}_K| } t^{-c} (\Zcal^i_{K+1} \mid \sbf_{K+1}, \Ycal^{j,i})
\\
    &\prod_{\xbf \in \Ucal} p_{\text{U}}(\xbf,\boldsymbol{s}_{K+1})  \prod_{i=1}^{|\Zcal^y_{K+1}|} \tilde{l}(\zbf^{y,i}_{K+1} \mid \sbf_{K+1}, \Ycal_i).
    \end{split}
\end{equation}

Putting these equations together, we get 
\begin{equation}
    \begin{split}
    & g(\mathcal{Z}_{1:K+1} \mid \boldsymbol{s}_{1:K+1},\mathcal{X}) = g(\mathcal{Z}_{1:K} \mid \boldsymbol{s}_{1:K},\mathcal{X})
g(\mathcal{Z}_{K+1} \mid \boldsymbol{s}_{K+1},\mathcal{X}) \label{eq:LikelihoodInduction1} \\
& = e^{-K\int c(\boldsymbol{z}) \text{d} \boldsymbol{z}}  \sum_{j\in \mathbb{J}_K}
\sum_{\mathcal{X}_{\mathrm{U}}\biguplus\mathcal{Y}^{j,1}\biguplus\dots\biguplus\mathcal{Y}^{j,|\mathbb{I}^{j}_K|}=\mathcal{X}}
\prod_{\boldsymbol{x} \in \mathcal{X}_{\mathrm{U}}}p_{\text{U}}(\boldsymbol{x},\boldsymbol{s}_{1:K}) 
\\ &  
\prod_{i=1}^{|\mathbb{I}^{j}_K|}t(\mathcal{Z}_{1:K}^{j,i}|\boldsymbol{s}_{1:K},\mathcal{Y}^{j,i}) e^{-\int c(\boldsymbol{z}) \text{d} \boldsymbol{z}} \sum_{\Zcal^1_{K+1} \biguplus \dots \biguplus Z^{ |\mathbb{I}^{j}_K|}_{K+1} \biguplus \Zcal^y_{K+1} = \Zcal_{K+1} }\\
    &
    \sum_{ \Ucal \biguplus \Ycal_1 \biguplus \dots \biguplus \Ycal_{ |\Zcal^y_{K+1}|} = \Xcal_\mathrm{U}  } \prod_{i=1}^{ |\mathbb{I}^{j}_K| } t^{-c} (\Zcal^i_{K+1} \mid \sbf_{K+1}, \Ycal^{j,i})\\
    & 
 \prod_{\xbf \in \Ucal} p_{\text{U}}(\xbf,\boldsymbol{s}_{K+1}) \prod_{i=1}^{|\Zcal^y_{K+1}|} \tilde{l}(\zbf^{y,i}_{K+1} \mid \sbf_{K+1}, \Ycal_i).
    \end{split}
\end{equation}
We can now reorder the summations and factors in \eqref{eq:LikelihoodInduction1} to resemble \eqref{eq:jointlikelihood}:
\begin{equation}
    \begin{split}
    & g(\mathcal{Z}_{1:K+1} \mid \boldsymbol{s}_{1:K+1},\mathcal{X})   \\
& = e^{-(K+1)\int c(\boldsymbol{z}) \text{d} \boldsymbol{z}}  \sum_{j\in \mathbb{J}_K} \sum_{\Zcal^1_{K+1} \biguplus \dots \biguplus Z^{ |\mathbb{I}^{j}_K|}_{K+1} \biguplus \Zcal^y_{K+1} = \Zcal_{K+1} } \\
&
\sum_{\mathcal{X}_{\mathrm{U}}\biguplus\mathcal{Y}^{j,1}\biguplus\dots\biguplus\mathcal{Y}^{j,|\mathbb{I}^{j}_K|}=\mathcal{X}} \sum_{ \Ucal \biguplus \Ycal_1 \biguplus \dots \biguplus \Ycal_{|\Zcal^y_{K+1}|} = \Xcal_\mathrm{U}  } 
\prod_{\boldsymbol{x} \in \mathcal{X}_{\mathrm{U}}}p_{\text{U}}(\boldsymbol{x},\boldsymbol{s}_{1:K}) 
\\ &  
\prod_{\xbf \in \Ucal} p_{\text{U}}(\xbf,\boldsymbol{s}_{K+1}) \prod_{i=1}^{|\mathbb{I}^{j}_K|}t(\mathcal{Z}_{1:K}^{j,i}|\boldsymbol{s}_{1:K},\mathcal{Y}^{j,i}) \\
&
 \prod_{i=1}^{ |\mathbb{I}^{j}_K| } t^{-c} (\Zcal^i_{K+1} \mid \sbf_{K+1}, \Ycal^{j,i})  
     \prod_{i=1}^{|\Zcal^y_{K+1}|} \tilde{l}(\zbf^{y,i}_{K+1} \mid \sbf_{K+1}, \Ycal_i).\label{eq:LikelihoodInduction3}
    \end{split}
\end{equation}
To further simplify the expression we note that 
\begin{equation}
    \prod_{\xbf \in \Xcal_{\mathrm{U}}}p_{\mathrm{U}}(\xbf,\sbf_{1:K}) = \prod_{\xbf \in \Ucal}p_{\text{U}}(\xbf,\sbf_{1:K}) \prod_{i=1}^{|\Zcal^y_{K+1}|} \prod_{\xbf \in \Ycal_i} p_{\text{U}}(\xbf,\sbf_{1:K}).
\end{equation}

We can remove $\Xcal_\mathrm{U}$ from the expression by merging the third and fourth summations in \eqref{eq:LikelihoodInduction3}:
\begin{equation}
    \begin{split}
    & g(\mathcal{Z}_{1:K+1} \mid \boldsymbol{s}_{1:K+1},\mathcal{X})    \\
& = e^{-(K+1)\int c(\boldsymbol{z}) \text{d} \boldsymbol{z}}  \sum_{j\in \mathbb{J}_K} \sum_{\Zcal^1_{K+1} \biguplus \dots \biguplus Z^{ |\mathbb{I}^{j}_K|}_{K+1} \biguplus \Zcal^y_{K+1} = \Zcal_{K+1} } \\
&
\sum_{ \Ucal \biguplus \Ycal_1 \biguplus \dots \biguplus \Ycal_{|\Zcal^y_{K+1}|} \biguplus\mathcal{Y}^{j,1}\biguplus\dots\biguplus\mathcal{Y}^{j,|\mathbb{I}^{j}_K|}=\mathcal{X}} 
\prod_{\xbf \in \Ucal} p_{\text{U}}(\xbf,\boldsymbol{s}_{1:K+1}) \\ & \prod_{i=1}^{|\mathbb{I}^{j}_K|}t(\mathcal{Z}_{1:K}^{j,i}|\boldsymbol{s}_{1:K},\mathcal{Y}^{j,i}) t^{-c} (\Zcal^i_{K+1} \mid \sbf_{K+1}, \Ycal^{j,i}) \\
& 
     \prod_{i=1}^{|\Zcal^y_{K+1}|} \left( \tilde{l}(\zbf^{y,i}_{K+1} \mid \sbf_{K+1}, \Ycal_i)\prod_{\xbf \in \Ycal_i} p_{\text{U}}(\xbf,\sbf_{1:K}) \right) 
     \\
& = e^{-(K+1)\int c(\boldsymbol{z}) \text{d} \boldsymbol{z}}  \sum_{j\in \mathbb{J}_K} \sum_{\Zcal^1_{K+1} \biguplus \dots \biguplus Z^{ |\mathbb{I}^{j}_K|}_{K+1} \biguplus \Zcal^y_{K+1} = \Zcal_{K+1} } \\
&
\sum_{ \Ucal \biguplus \Ycal_1 \biguplus \dots \biguplus \Ycal_{|\Zcal^y_{K+1}|} \biguplus\mathcal{Y}^{j,1}\biguplus\dots\biguplus\mathcal{Y}^{j,|\mathbb{I}^{j}_K|}=\mathcal{X}} 
\\ &  
\prod_{\xbf \in \Ucal} p_{\text{U}}(\xbf,\boldsymbol{s}_{1:K+1}) \prod_{i=1}^{|\mathbb{I}^{j}_K|}t((\mathcal{Z}_{1:K}^{j,i}, \Zcal^i_{K+1} )|\boldsymbol{s}_{1:K+1},\mathcal{Y}^{j,i})  \\
& 
     \prod_{i=1}^{|\Zcal^y_{K+1}|} t( (\emptyset,\dots,\emptyset, \{\zbf^{y,i}_{K+1}\}) \mid \sbf_{1:K+1}, \Ycal_i).\label{eq:LikelihoodInduction4} 
    \end{split}
\end{equation}

The first two summations, over $j\in \mathbb{J}_K$ and $\Zcal^1_{K+1} \biguplus \dots \biguplus Z^{ |\mathbb{I}^{j}_K|}_{K+1} \biguplus \Zcal^y_{K+1} = \Zcal_{K+1}$, sum over all valid partitions of $\Zcal_{1:K+1}$. We use $\mathbb{J}_{K+1}$ to denote the index set of all valid partitions of $\Zcal_{1:K+1}$, therefore, the first two summations are equivalent to a sum over all $j \in \mathbb{J}_{K+1}$. More specifically, for every $j \in \mathbb{J}_{K+1}$ there is precisely one term in this double summation. The double summation may also contain some illegal partitions, e.g., $|\Zcal^1_{K+1}|>1$, but these do not contribute to the sum since $t^{-c} (\Zcal^i_{K+1} \mid \sbf_{K+1}, \Ycal^{j,i})=0$ for those partitions.   

The third summation instead specifies which sets of landmarks gave rise to the different subsets in the partition: $\Ucal$ are landmarks that are undetected at all time steps, $\Ycal_i$ are (possibly empty) sets of newly detected landmarks (they are detected for the first time at time step $K+1$), and $\Ycal^{j,i}$ are (possibly empty) sets of landmarks detected before time step $K+1$. As for the factors inside the summations, $e^{-(K+1)\int c(\boldsymbol{z}) \text{d} \boldsymbol{z}}$ and $\prod_{\xbf \in \Ucal} p_{\text{U}}(\xbf,\boldsymbol{s}_{1:K+1})$ match \eqref{eq:jointlikelihood}. We also note that $(\mathcal{Z}_{1:K}^{j,i},\Zcal^i_{K+1})$ jointly define the sequence of measurements generated by $\Ycal^{j,i}$, and that $(\emptyset,\dots,\emptyset, \{\zbf^{y,i}_{K+1}\})$ define the sequence of measurements generated by $\Ycal_i$.

To simplify \eqref{eq:LikelihoodInduction4}, we re-index all detected landmarks $\Ycal_1 \biguplus \dots \biguplus \Ycal_{|\Zcal^y_{K+1}|} \biguplus\mathcal{Y}^{j,1}\biguplus\dots\biguplus\mathcal{Y}^{j,|\mathbb{I}^{j}_K|}$ with $ \mathbb{I}^{j}_{K+1}=\{1,\cdots,|\mathbb{I}^{j}_{K+1}|\}$, where $ |\mathbb{I}^{j}_{K+1}|=|\mathbb{I}^{j}_{K}|+|\Zcal^y_{K+1}|$, resulting in $\mathcal{Y}^{j,1}\biguplus\dots\biguplus\mathcal{Y}^{j,|\mathbb{I}^{j}_{K+1}|}$. The first $|\Zcal^y_{K+1}|$ sets correspond to the newly detected landmarks, i.e., $\Ycal_1,\dots, \Ycal_{|\Zcal^y_{K+1}|}$, and the last $|\mathbb{I}^{j}_K|$ sets correspond to landmarks detected before time step $K+1$, i.e., $\mathcal{Y}^{j,1},\dots,\mathcal{Y}^{j,|\mathbb{I}^{j}_K|}$. In addition, we use $\mathcal{Z}_{1:K+1}^{j,i}$ to denote the sequence of measurements generated from $\mathcal{Y}^{j,i}$, where $\mathcal{Z}_{1:K+1}^{j,i}=(\emptyset,\dots,\emptyset, \{\zbf^{y,i}_{K+1}\}), \forall i \in \{1,\cdots,|\Zcal^y_{K+1}|\}$, and $\mathcal{Z}_{1:K+1}^{j,i}=(\mathcal{Z}_{1:K}^{j,i},\Zcal^i_{K+1}), \forall i \in \{|\Zcal^y_{K+1}|+1,\cdots,|\mathbb{I}^{j}_{K+1}|\}$ for the sequences of measurements generated from newly detected and previously detected landmarks, respectively. We therefore merge the last two products, and replace the detected landmarks with $\mathcal{Y}^{j,1}\biguplus\dots\biguplus\mathcal{Y}^{j,|\mathbb{I}^{j}_{K+1}|}$ in \eqref{eq:LikelihoodInduction4}, resulting in:
\begin{align}
   & g(\mathcal{Z}_{1:K+1} \mid \boldsymbol{s}_{1:K+1},\mathcal{X} ) = e^{-(K+1)\int c(\boldsymbol{z}) \text{d} \boldsymbol{z}} \label{eq:jointlikelihood_nextstep}
\\ & \sum_{j\in \mathbb{J}_K} \sum_{\Zcal^1_{K+1} \biguplus \dots \biguplus Z^{ |\mathbb{I}^{j}_K|}_{K+1} \biguplus \Zcal^y_{K+1} = \Zcal_{K+1} } 
\sum_{\Ucal\biguplus\mathcal{Y}^{j,1}\biguplus\dots\biguplus\mathcal{Y}^{j,|\mathbb{I}^{j}_{K+1}|}=\mathcal{X}} \nonumber\\
&
\prod_{\boldsymbol{x} \in \Ucal}p_{\text{U}}(\boldsymbol{x},\boldsymbol{s}_{1:K+1})  
\prod_{i=1}^{|\mathbb{I}^{j}_{K+1}|}t(\mathcal{Z}_{1:K+1}^{j,i}|\boldsymbol{s}_{1:K+1},\mathcal{Y}^{j,i}).\nonumber
\end{align}
To further simplify the expression, we replace the first two summations with a sum over all $j \in \mathbb{J}_{K+1}$. For each new $j$, there is precisely one case in the union of all detected landmarks, and we denote $\tilde{\mathbb{I}}^{j}_{K+1}$ as the index set of all detected landmarks for the $j$-th new partition. Then, we have
\begin{align}
   & g(\mathcal{Z}_{1:K+1} \mid \boldsymbol{s}_{1:K+1},\mathcal{X} ) = e^{-(K+1)\int c(\boldsymbol{z}) \text{d} \boldsymbol{z}} \label{eq:jointlikelihood_nextstep1}
\\ & \sum_{j\in \mathbb{J}_{K+1}}
\sum_{\Ucal\biguplus\mathcal{Y}^{j,1}\biguplus\dots\biguplus\mathcal{Y}^{j,|\tilde{\mathbb{I}}^{j}_{K+1}|}=\mathcal{X}}
\prod_{\boldsymbol{x} \in \Ucal}p_{\text{U}}(\boldsymbol{x},\boldsymbol{s}_{1:K+1}) \nonumber
\\ &  
\prod_{i=1}^{|\tilde{\mathbb{I}}^{j}_{K+1}|}t(\mathcal{Z}_{1:K+1}^{j,i}|\boldsymbol{s}_{1:K+1},\mathcal{Y}^{j,i}).\nonumber
\end{align}

Since \eqref{eq:jointlikelihood_nextstep1} matches \eqref{eq:jointlikelihood}, this confirms that \eqref{eq:jointlikelihood} holds for $K+1$. Thus, we have completed the proof of Lemma \ref{lemma:BatchLikelihood}.

\end{proof}

Theorem \ref{theo:posterior} is then proved by plugging \eqref{eq:jointlikelihood} 
 and the expression for the \ac{PPP} prior in \eqref{PPP} into \eqref{jointposter_mark}.

\section{Gradient Descent in GraphSLAM} \label{sec:gradient}

This appendix explains how to apply gradient descent to solve \eqref{eq:optimizationProblem_min}. We  minimize the increments $\Delta \hat{\boldsymbol{q}}$, denoted as
\begin{align}
 &\Delta \hat{\boldsymbol{q}}=\arg \underset{\Delta \boldsymbol{q}} {\min}  ~ {\mathcal{E}}(\hat{\boldsymbol{q}}+\Delta \boldsymbol{q}).\label{eq:optimizationProblem_increment}  
\end{align}
To achieve this, we firstly replace the nonlinear components in ${\mathcal{E}}(\hat{\boldsymbol{q}}+\Delta \boldsymbol{q})$ with their approximations, i.e. 
\begin{align}
 &\boldsymbol{v}(\boldsymbol{\epsilon}_{k}+\Delta \boldsymbol{s}_{k}) \approx \boldsymbol{v}(\boldsymbol{\epsilon}_{k}) + \boldsymbol{F}_{k}\Delta \boldsymbol{s}_{k},\label{eq:linear_motion}  \\&
 \boldsymbol{h}(\hat{\boldsymbol{q}}^{i}_{k}+\Delta\boldsymbol{q}^{i}_{k})\approx\boldsymbol{h}(\hat{\boldsymbol{q}}^{i}_{k})+\boldsymbol{H}^{i}_{k}\Delta\boldsymbol{q}^{i}_{k},\label{eq:linear_measurement}
\end{align}
where the matrix $\boldsymbol{F}_{k}$ denotes the Jacobian of $\boldsymbol{v}(\boldsymbol{s}_{k})$, evaluated at $\boldsymbol{s}_{k}=\boldsymbol{\epsilon}_{k}$, i.e., $   \boldsymbol{F}_{k}=\left. {\partial \boldsymbol{v}(\boldsymbol{s}_{k})}/{\partial \boldsymbol{s}_{k}}\right|_{\boldsymbol{s}_{k}=\boldsymbol{\epsilon}_{k}}$, 
the matrix $\boldsymbol{H}^{i}_{k}$ denotes the Jacobian of $\boldsymbol{h}(\boldsymbol{q}^{i}_{k})$, evaluated at $\boldsymbol{q}^{i}_{k}=\hat{\boldsymbol{q}}^{i}_{k}$, i.e., $\boldsymbol{H}^{i}_{k} =\left. {\partial \boldsymbol{h}(\boldsymbol{q}^{i}_{k})}/{\partial \boldsymbol{q}^{i}_{k}}\right|_{\boldsymbol{q}^{i}_{k}=\hat{\boldsymbol{q}}^{i}_{k}},$ and it can be decomposed as $\boldsymbol{H}^{i}_{k}=[(\boldsymbol{H}^{i}_{k,\text{S}})^{\textsf{T}},(\boldsymbol{H}^{i}_{k,\text{L}})^{\textsf{T}}]^{\textsf{T}}$, compromising the state part $\boldsymbol{H}^{i}_{k,\text{S}}$ and the landmark part $\boldsymbol{H}^{i}_{k,\text{L}}$.

By expanding ${\mathcal{E}}(\hat{\boldsymbol{q}}+\Delta \boldsymbol{q})$, putting the same items together and concatenating individual components with the same sequence of $\boldsymbol{q}$, we can have
\begin{align}
 {\mathcal{E}}(\hat{\boldsymbol{q}}+& \Delta \boldsymbol{q})\approx  
 e +  \boldsymbol{b}^{\textsf{T}} \Delta \boldsymbol{q}+ (\Delta \boldsymbol{q})^{\textsf{T}}\boldsymbol{b} +  (\Delta\boldsymbol{q})^{\textsf{T}}\boldsymbol{\Omega}\Delta\boldsymbol{q}\label{eq:optimizationProblem_min_final}, 
\end{align}
where $e$ is a constant, $\boldsymbol{b}$ comprises the specific error terms of the corresponding components, which is a vector with the size $d_{\boldsymbol{q}} \times 1$, and $\boldsymbol{\Omega}$ is the information matrix with the size $d_{\boldsymbol{q}} \times d_{\boldsymbol{q}}$. How to construct  $\boldsymbol{b}$ and $\boldsymbol{\Omega}$ can be found, e.g., in \cite[Section 5.4]{thrun2006graph}. The minimum of the quadratic approximating \eqref{eq:optimizationProblem_min_final} around $\boldsymbol{q}$ admits a  closed-form solution 
\begin{align}
 &\Delta \hat{\boldsymbol{q}}=-\boldsymbol{\Omega}^{-1}\boldsymbol{b},\label{best_increment}
\end{align}
and the updated state can be given by
\begin{align}
 &\hat{\boldsymbol{q}}\gets \hat{\boldsymbol{q}}+\Delta \hat{\boldsymbol{q}}.\label{updadted_state}
\end{align}
The information matrix of $\boldsymbol{q}$ is $\boldsymbol{\Omega}$, so that the covariance of $\boldsymbol{q}$ is the inverse of $\boldsymbol{\Omega}$, denoted as $\boldsymbol{\Omega}^{-1}$. Then, the updated joint state follows the Gaussian distribution  $\mathcal{N}(\boldsymbol{q};\hat{\boldsymbol{q}},\boldsymbol{\Omega}^{-1})$. We repeat \eqref{eq:optimizationProblem_increment} to \eqref{updadted_state} until it converges or reaches the maximum number of iterations. 